\documentclass{article}

\usepackage{arxiv}

\usepackage[utf8]{inputenc} 
\usepackage[T1]{fontenc}    
\usepackage{hyperref}       
\usepackage{url}            
\usepackage{booktabs}       
\usepackage{amsfonts}       
\usepackage{nicefrac}       
\usepackage{microtype}      
\usepackage{lipsum}
\usepackage{graphicx}

\title{Hospital Capacity Planning Using Discrete Event Simulation Under
Special Consideration of the COVID-19 Pandemic}

\author{
    Thomas Bartz-Beielstein
   \\
    IDEA, TH Köln \\
  Steinmüllerallee 1, 51643 Gummersbach, Germany \\
  \texttt{\href{mailto:thomas.bartz-beielstein@th-koeln.de}{\nolinkurl{thomas.bartz-beielstein@th-koeln.de}}} \\
   \And
    Frederik Rehbach
   \\
    IDEA, TH Köln \\
  Steinmüllerallee 1, 51643 Gummersbach, Germany \\
  \texttt{\href{mailto:frederik.rehbach@th-koeln.de}{\nolinkurl{frederik.rehbach@th-koeln.de}}} \\
   \And
    Olaf Mersmann
   \\
    IDEA, TH Köln \\
  Steinmüllerallee 1, 51643 Gummersbach, Germany \\
  \texttt{\href{mailto:olaf.mersmann@th-koeln.de}{\nolinkurl{olaf.mersmann@th-koeln.de}}} \\
   \And
    Eva Bartz
   \\
    Bartz und Bartz GmbH \\
  Goebenstr. 10, 51643 Gummersbach, Germany \\
  \texttt{\href{mailto:eva.bartz@bartzundbartz.de}{\nolinkurl{eva.bartz@bartzundbartz.de}}} \\
  }

\usepackage{color}
\usepackage{fancyvrb}

\DefineVerbatimEnvironment{Highlighting}{Verbatim}{commandchars=\\\{\}}
\usepackage{framed}
\definecolor{shadecolor}{RGB}{248,248,248}
\newenvironment{Shaded}{\begin{snugshade}}{\end{snugshade}}

\newcommand{\AttributeTok}[1]{\textcolor[rgb]{0.77,0.63,0.00}{#1}}

\newcommand{\ConstantTok}[1]{\textcolor[rgb]{0.00,0.00,0.00}{#1}}

\newcommand{\DecValTok}[1]{\textcolor[rgb]{0.00,0.00,0.81}{#1}}

\newcommand{\ErrorTok}[1]{\textcolor[rgb]{0.64,0.00,0.00}{\textbf{#1}}}

\newcommand{\FloatTok}[1]{\textcolor[rgb]{0.00,0.00,0.81}{#1}}
\newcommand{\FunctionTok}[1]{\textcolor[rgb]{0.00,0.00,0.00}{#1}}

\newcommand{\NormalTok}[1]{#1}

\newcommand{\OtherTok}[1]{\textcolor[rgb]{0.56,0.35,0.01}{#1}}

\newcommand{\SpecialCharTok}[1]{\textcolor[rgb]{0.00,0.00,0.00}{#1}}

\newcommand{\StringTok}[1]{\textcolor[rgb]{0.31,0.60,0.02}{#1}}

\newlength{\csllabelwidth}
\newlength{\cslhangindent}
  {}%
  {\par}
\newenvironment{CSLReferences}[3] 
 {
  \setlength{\parindent}{0pt}
  \ifodd #1 \everypar{\setlength{\hangindent}{\cslhangindent}}\ignorespaces\fi
  \ifnum #2 > 0
  \setlength{\parskip}{#3\baselineskip}
  \fi
 }%
 {}
\usepackage{calc} 

\begin{document}
\maketitle

\def\tightlist{}

\begin{abstract}
We present a resource-planning tool for hospitals under special
consideration of the COVID-19 pandemic, called babsim.hospital. It
provides many advantages for crisis teams, e.g., comparison with their
own local planning, simulation of local events, simulation of several
scenarios (worst / best case). There are benefits for medical
professionals, e.g, analysis of the pandemic at local, regional, state
and federal level, the consideration of special risk groups, tools for
validating the length of stays and transition probabilities. Finally,
there are potential advantages for administration, management, e.g.,
assessment of the situation of individual hospitals taking local events
into account, consideration of relevant resources such as beds,
ventilators, rooms, protective clothing, and personnel planning, e.g.,
medical and nursing staff. \texttt{babsim.hospital} combines simulation,
optimization, statistics, and artificial intelligence processes in a
very efficient way. The core is a discrete, event-based simulation
model.
\end{abstract}

\keywords{
    Statistics
   \and
    Applications, Computer Science
   \and
    Computers and Society
  }

\hypertarget{introduction}{%
\section{Introduction}\label{introduction}}

Resource and capacity planning with respect to the COVID-19 pandemic is
a challenging task for hospitals. This paper describes
\texttt{BaBSimHospital}, a resource planning tool that was developed in
cooperation with ICU experts, crisis teams, and health administrations.
\texttt{babsim.hospital} combines simulation, optimization, statistics,
and artificial intelligence processes in a very efficient way. The core
is a discrete, event-based simulation model. Sequential parameter
optimization (SPOT) is used to optimize the parameters (status
transition probabilities, length of stay, distribution properties)
Bartz-Beielstein, Lasarczyk, and Preuss (2005). For modeling purposes,
distributions (including a truncated and translated Gamma distribution)
were specially developed by us in order to realistically simulate the
length of stay. babsim.hospital takes into account different risks for
individual groups of people (age and gender-specific) and can be used
for any resources beyond the planning of bed capacities.

Using \texttt{BaBSimHospital} provides the following advantages for
crisis teams:

\begin{itemize}
\item comparison with your own local planning,
\item simulation of local events,
\item adaptation to your own situation,
\item simulation of any scenario (worst / best case),
\item simulation of any pandemic scenario, considering the local situation,
\item standardized approach.
\end{itemize}

And, there are benefits for medical professionals, e.g,

\begin{itemize}
\item analysis of the pandemic at local, regional, state and federal level,
\item consideration of special risk groups,
\item validation of the length of stay,
\item validation of the probabilities.
\end{itemize}

Finally, there are potential advantages for administration, management,
e.g.,

\begin{itemize}
\item assessment of the situation of individual hospitals taking local events into account,
\item consideration of relevant resources: beds, ventilators, rooms, protective clothing,
\item personnel planning: medical and nursing staff.
\end{itemize}

The ideas used in the \texttt{babsim.hospital} implementation are based
on the paper by Lawton and McCooe (2019). \texttt{babsim.hospital} is
implemented in the statistical programming language R, see R Core Team
(2020). It uses a discrete-event simulation, which uses the
\texttt{simmer} package (Ucar, Smeets, and Azcorra 2019).

This paper describes practical aspects (``how to use'') of the
\texttt{babsim.hospital} package. Although this paper focuses on the
situation in Germany, it can be used for different settings. For
example, we have used data from UK to model a local configuration. Even,
if there are no real-world data available, the user can perform
simulations based on synthetic data. How to use synthetic data and the
theoretical background of \texttt{babsim.hospital} is described in
Bartz-Beielstein et al. (2020).

This paper is structured as follows. Section\ref{sec:data} describes the
data, which are necessary for the simulation. How to run a simulation is
described in Section \ref{sec:simulation}. The visualization of
simulation results is illustrated in Section \ref{sec:visualization}.
Section \ref{sec:visualization} also presents the error measure, which
is used for the optimizations. The optimization procedure is introduced
in Section \ref{sec:optimization}. How to visualize the optimized
parameter settings is discussed in Section \ref{sec:vispara}. Section
\ref{sec:extend} describes how data for future simulations can be
generated. \texttt{babsim.hospital} provides tools for a deeper
understanding of the simulation parameters. The corresponding tools from
statistics and sensitivity analysis are described in Section
\ref{sec:sensitivity}.

\hypertarget{packages-and-data}{%
\section{Packages and Data}\label{packages-and-data}}

\hypertarget{packages}{%
\subsection{Packages}\label{packages}}

A clean start and loading the required packages can be performed as
follows:

\begin{Shaded}
\begin{Highlighting}[]
\FunctionTok{rm}\NormalTok{(}\AttributeTok{list=}\FunctionTok{ls}\NormalTok{())}
\FunctionTok{suppressPackageStartupMessages}\NormalTok{(\{}
\FunctionTok{library}\NormalTok{(}\StringTok{"SPOT"}\NormalTok{)}
\FunctionTok{library}\NormalTok{(}\StringTok{"babsim.hospital"}\NormalTok{)}
\FunctionTok{library}\NormalTok{(}\StringTok{"simmer"}\NormalTok{)}
\FunctionTok{library}\NormalTok{(}\StringTok{"simmer.plot"}\NormalTok{)}
\FunctionTok{library}\NormalTok{(}\StringTok{"plotly"}\NormalTok{)}
\FunctionTok{library}\NormalTok{(}\StringTok{"rpart"}\NormalTok{)}
\FunctionTok{library}\NormalTok{(}\StringTok{"rpart.plot"}\NormalTok{)}
\NormalTok{\})}
\end{Highlighting}
\end{Shaded}

We need at least version 2.1.8 of \texttt{SPOT}.

\begin{Shaded}
\begin{Highlighting}[]
\FunctionTok{packageVersion}\NormalTok{(}\StringTok{"SPOT"}\NormalTok{)}
\NormalTok{\%}\SpecialCharTok{\textgreater{}}\NormalTok{ [}\DecValTok{1}\NormalTok{] }\StringTok{\textquotesingle{}2.1.10\textquotesingle{}}
\end{Highlighting}
\end{Shaded}

\hypertarget{simulation-and-field-data}{%
\subsection{Simulation and Field Data}\label{simulation-and-field-data}}

\label{sec:data}

The \texttt{babsim.hospital} simulator models resources usage in
hospitals, e.g., number of ICU beds (\(y\)), as a function of the number
of infected individuals (\(x\)). In addition to the number of
infections, information about age and gender will be used as simulation
input.

In general, the simulator requires two types of data:
\emph{simulation data} that describes the spread of the pandemic over
time and \emph{field data} that contains daily resource usage data.
Included in the package are tools to generate synthetic data, e.g., you
can generate simulation and field data to run the simulations. This
procedure is described in Bartz-Beielstein et al. (2020).

To demonstrate the usage of real-world data, we have included two sample
datasets from Germany.

\begin{itemize}
\item Simulation data, i.e., input data for the simulation. We have included a data sample from the German [Robert Koch-Institute](https://www.rki.de) (RKI).
Please take the copyright notice, 
which is shown in Section \ref{sec:appendix} (Appendix) under advisement, if you plan to use the RKI data included in the package.
\item Field data. We have included a data sample from the German [DIVI Register](https://www.intensivregister.de/). 
The field data is used to validate the output of the simulation.
Please take the copyright notice, 
which is shown in Section \ref{sec:appendix} (Appendix) under advisement, if you plan to use the RKI data included in the package.
\end{itemize}

\hypertarget{simulation-data}{%
\subsubsection{Simulation Data}\label{simulation-data}}

First, we will take a closer look at the simulation data. Here, we will
use the data from the RKI Server. \texttt{babsim.hospital} provides a
function to update the (daily) RKI data.

\begin{Shaded}
\begin{Highlighting}[]
\FunctionTok{updateRkidataFile}\NormalTok{(}\StringTok{"https://www.arcgis.com/sharing/rest/content/items/f10774f1c63e40168479a1feb6c7ca74/data"}\NormalTok{)}
\end{Highlighting}
\end{Shaded}

Users are expected to adapt this function to their local situation. The
downloaded data will be available in the package as \texttt{rkidata}.
Due to data size limits on CRAN, the full dataset is not included in the
\texttt{babsim.hospital}package. Instead, we provide a subset of the
Robert Koch-Institut dataset with 10,000 observations in the package.

\begin{Shaded}
\begin{Highlighting}[]
\FunctionTok{str}\NormalTok{(babsim.hospital}\SpecialCharTok{::}\NormalTok{rkidata)}
\NormalTok{\%}\SpecialCharTok{\textgreater{}} \StringTok{\textquotesingle{}data.frame\textquotesingle{}}\SpecialCharTok{:}    \DecValTok{500734}\NormalTok{ obs. of  }\DecValTok{18}\NormalTok{ variables}\SpecialCharTok{:}
\NormalTok{\%}\SpecialCharTok{\textgreater{}}  \ErrorTok{$}\NormalTok{ FID                 }\SpecialCharTok{:}\NormalTok{ int  }\DecValTok{32} \DecValTok{86} \DecValTok{87} \DecValTok{88} \DecValTok{89} \DecValTok{90} \DecValTok{91} \DecValTok{92} \DecValTok{93} \DecValTok{94}\NormalTok{ ...}
\NormalTok{\%}\SpecialCharTok{\textgreater{}}  \ErrorTok{$}\NormalTok{ IdBundesland        }\SpecialCharTok{:}\NormalTok{ int  }\DecValTok{1} \DecValTok{1} \DecValTok{1} \DecValTok{1} \DecValTok{1} \DecValTok{1} \DecValTok{1} \DecValTok{1} \DecValTok{1} \DecValTok{1}\NormalTok{ ...}
\NormalTok{\%}\SpecialCharTok{\textgreater{}}  \ErrorTok{$}\NormalTok{ Bundesland          }\SpecialCharTok{:}\NormalTok{ chr  }\StringTok{"Schleswig{-}Holstein"} \StringTok{"Schleswig{-}Holstein"} \StringTok{"Schleswig{-}Holstein"} \StringTok{"Schleswig{-}Holstein"}\NormalTok{ ...}
\NormalTok{\%}\SpecialCharTok{\textgreater{}}  \ErrorTok{$}\NormalTok{ Landkreis           }\SpecialCharTok{:}\NormalTok{ chr  }\StringTok{"SK Kiel"} \StringTok{"SK Kiel"} \StringTok{"SK Kiel"} \StringTok{"SK Kiel"}\NormalTok{ ...}
\NormalTok{\%}\SpecialCharTok{\textgreater{}}  \ErrorTok{$}\NormalTok{ Altersgruppe        }\SpecialCharTok{:}\NormalTok{ chr  }\StringTok{"A15{-}A34"} \StringTok{"A35{-}A59"} \StringTok{"A35{-}A59"} \StringTok{"A35{-}A59"}\NormalTok{ ...}
\NormalTok{\%}\SpecialCharTok{\textgreater{}}  \ErrorTok{$}\NormalTok{ Geschlecht          }\SpecialCharTok{:}\NormalTok{ chr  }\StringTok{"M"} \StringTok{"M"} \StringTok{"M"} \StringTok{"M"}\NormalTok{ ...}
\NormalTok{\%}\SpecialCharTok{\textgreater{}}  \ErrorTok{$}\NormalTok{ AnzahlFall          }\SpecialCharTok{:}\NormalTok{ int  }\DecValTok{1} \DecValTok{1} \DecValTok{1} \DecValTok{1} \DecValTok{1} \DecValTok{1} \DecValTok{1} \DecValTok{1} \DecValTok{1} \DecValTok{1}\NormalTok{ ...}
\NormalTok{\%}\SpecialCharTok{\textgreater{}}  \ErrorTok{$}\NormalTok{ AnzahlTodesfall     }\SpecialCharTok{:}\NormalTok{ int  }\DecValTok{0} \DecValTok{0} \DecValTok{0} \DecValTok{0} \DecValTok{0} \DecValTok{0} \DecValTok{0} \DecValTok{0} \DecValTok{0} \DecValTok{0}\NormalTok{ ...}
\NormalTok{\%}\SpecialCharTok{\textgreater{}}  \ErrorTok{$}\NormalTok{ Refdatum            }\SpecialCharTok{:}\NormalTok{ chr  }\StringTok{"2020/09/01 00:00:00"} \StringTok{"2020/09/02 00:00:00"} \StringTok{"2020/09/03 00:00:00"} \StringTok{"2020/09/08 00:00:00"}\NormalTok{ ...}
\NormalTok{\%}\SpecialCharTok{\textgreater{}}  \ErrorTok{$}\NormalTok{ IdLandkreis         }\SpecialCharTok{:}\NormalTok{ int  }\DecValTok{1002} \DecValTok{1002} \DecValTok{1002} \DecValTok{1002} \DecValTok{1002} \DecValTok{1002} \DecValTok{1002} \DecValTok{1002} \DecValTok{1002} \DecValTok{1002}\NormalTok{ ...}
\NormalTok{\%}\SpecialCharTok{\textgreater{}}  \ErrorTok{$}\NormalTok{ Datenstand          }\SpecialCharTok{:}\NormalTok{ chr  }\StringTok{"12.12.2020, 00:00 Uhr"} \StringTok{"12.12.2020, 00:00 Uhr"} \StringTok{"12.12.2020, 00:00 Uhr"} \StringTok{"12.12.2020, 00:00 Uhr"}\NormalTok{ ...}
\NormalTok{\%}\SpecialCharTok{\textgreater{}}  \ErrorTok{$}\NormalTok{ NeuerFall           }\SpecialCharTok{:}\NormalTok{ int  }\DecValTok{0} \DecValTok{0} \DecValTok{0} \DecValTok{0} \DecValTok{0} \DecValTok{0} \DecValTok{0} \DecValTok{0} \DecValTok{0} \DecValTok{0}\NormalTok{ ...}
\NormalTok{\%}\SpecialCharTok{\textgreater{}}  \ErrorTok{$}\NormalTok{ NeuerTodesfall      }\SpecialCharTok{:}\NormalTok{ int  }\SpecialCharTok{{-}}\DecValTok{9} \SpecialCharTok{{-}}\DecValTok{9} \SpecialCharTok{{-}}\DecValTok{9} \SpecialCharTok{{-}}\DecValTok{9} \SpecialCharTok{{-}}\DecValTok{9} \SpecialCharTok{{-}}\DecValTok{9} \SpecialCharTok{{-}}\DecValTok{9} \SpecialCharTok{{-}}\DecValTok{9} \SpecialCharTok{{-}}\DecValTok{9} \SpecialCharTok{{-}}\DecValTok{9}\NormalTok{ ...}
\NormalTok{\%}\SpecialCharTok{\textgreater{}}  \ErrorTok{$}\NormalTok{ Meldedatum          }\SpecialCharTok{:}\NormalTok{ chr  }\StringTok{"2020/08/27 00:00:00"} \StringTok{"2020/08/30 00:00:00"} \StringTok{"2020/09/03 00:00:00"} \StringTok{"2020/09/05 00:00:00"}\NormalTok{ ...}
\NormalTok{\%}\SpecialCharTok{\textgreater{}}  \ErrorTok{$}\NormalTok{ NeuGenesen          }\SpecialCharTok{:}\NormalTok{ int  }\DecValTok{0} \DecValTok{0} \DecValTok{0} \DecValTok{0} \DecValTok{0} \DecValTok{0} \DecValTok{0} \DecValTok{0} \DecValTok{0} \DecValTok{0}\NormalTok{ ...}
\NormalTok{\%}\SpecialCharTok{\textgreater{}}  \ErrorTok{$}\NormalTok{ AnzahlGenesen       }\SpecialCharTok{:}\NormalTok{ int  }\DecValTok{1} \DecValTok{1} \DecValTok{1} \DecValTok{1} \DecValTok{1} \DecValTok{1} \DecValTok{1} \DecValTok{1} \DecValTok{1} \DecValTok{1}\NormalTok{ ...}
\NormalTok{\%}\SpecialCharTok{\textgreater{}}  \ErrorTok{$}\NormalTok{ IstErkrankungsbeginn}\SpecialCharTok{:}\NormalTok{ int  }\DecValTok{1} \DecValTok{1} \DecValTok{0} \DecValTok{1} \DecValTok{0} \DecValTok{1} \DecValTok{1} \DecValTok{1} \DecValTok{1} \DecValTok{0}\NormalTok{ ...}
\NormalTok{\%}\SpecialCharTok{\textgreater{}}  \ErrorTok{$}\NormalTok{ Altersgruppe2       }\SpecialCharTok{:}\NormalTok{ chr  }\StringTok{"Nicht übermittelt"} \StringTok{"Nicht übermittelt"} \StringTok{"Nicht übermittelt"} \StringTok{"Nicht übermittelt"}\NormalTok{ ...}
\end{Highlighting}
\end{Shaded}

The \texttt{rkidata} can be visualized as follows (here
\texttt{region\ =\ 0} is Germany, \texttt{region\ =\ 5} is North
Rhine-Westphalia, \texttt{region\ =\ 5374} Oberbergischer Kreis, etc.):

\begin{Shaded}
\begin{Highlighting}[]
\NormalTok{p }\OtherTok{\textless{}{-}} \FunctionTok{ggVisualizeRki}\NormalTok{(}\AttributeTok{data=}\NormalTok{babsim.hospital}\SpecialCharTok{::}\NormalTok{rkidata, }\AttributeTok{region =} \DecValTok{5374}\NormalTok{)}
\FunctionTok{print}\NormalTok{(p)}
\end{Highlighting}
\end{Shaded}

\includegraphics[width=1\linewidth]{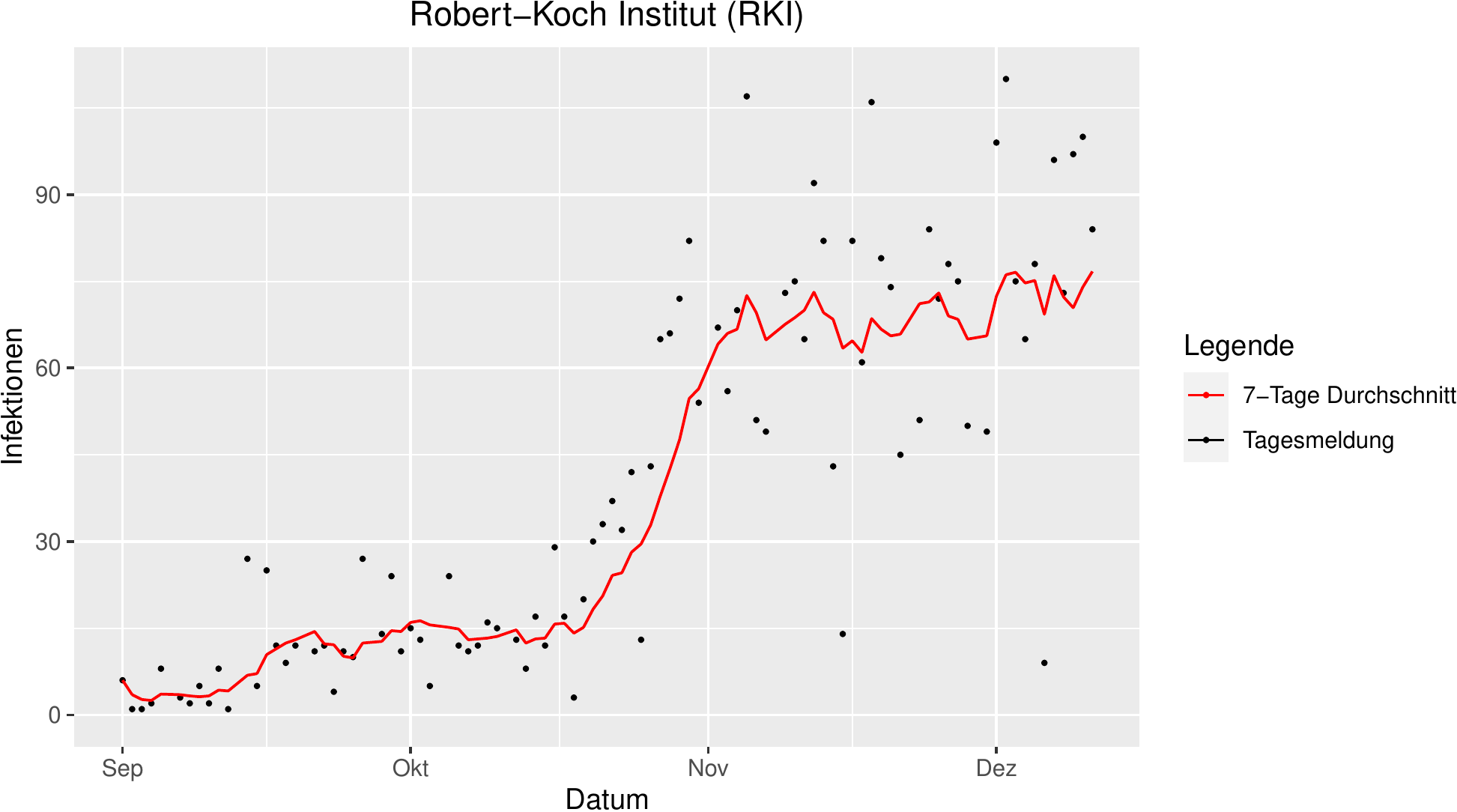}

After downloading the \texttt{rkidata}set, these data will be
preprocessed. Not all the information from the original \texttt{rkidata}
data set is required by the \texttt{babsim.hospital} simulator. The
function \texttt{getRkiData()} extracts the subset of the raw
\texttt{rkidata} required by our simulation, optimization, and analysis:

\begin{Shaded}
\begin{Highlighting}[]
\NormalTok{rki }\OtherTok{\textless{}{-}} \FunctionTok{getRkiData}\NormalTok{(}\AttributeTok{rki =}\NormalTok{ rkidata)}
\NormalTok{\%}\SpecialCharTok{\textgreater{}} \FunctionTok{getRkiData}\NormalTok{()}\SpecialCharTok{:}\NormalTok{ Found days with negative number of cases. Ignoring them.}
\FunctionTok{str}\NormalTok{(rki)}
\NormalTok{\%}\SpecialCharTok{\textgreater{}} \StringTok{\textquotesingle{}data.frame\textquotesingle{}}\SpecialCharTok{:}    \DecValTok{1055516}\NormalTok{ obs. of  }\DecValTok{7}\NormalTok{ variables}\SpecialCharTok{:}
\NormalTok{\%}\SpecialCharTok{\textgreater{}}  \ErrorTok{$}\NormalTok{ Altersgruppe}\SpecialCharTok{:}\NormalTok{ chr  }\StringTok{"A15{-}A34"} \StringTok{"A35{-}A59"} \StringTok{"A15{-}A34"} \StringTok{"A35{-}A59"}\NormalTok{ ...}
\NormalTok{\%}\SpecialCharTok{\textgreater{}}  \ErrorTok{$}\NormalTok{ Geschlecht  }\SpecialCharTok{:}\NormalTok{ chr  }\StringTok{"M"} \StringTok{"W"} \StringTok{"M"} \StringTok{"W"}\NormalTok{ ...}
\NormalTok{\%}\SpecialCharTok{\textgreater{}}  \ErrorTok{$}\NormalTok{ Day         }\SpecialCharTok{:}\NormalTok{ Date, format}\SpecialCharTok{:} \StringTok{"2020{-}09{-}01"} \StringTok{"2020{-}09{-}01"}\NormalTok{ ...}
\NormalTok{\%}\SpecialCharTok{\textgreater{}}  \ErrorTok{$}\NormalTok{ IdBundesland}\SpecialCharTok{:}\NormalTok{ int  }\DecValTok{1} \DecValTok{1} \DecValTok{1} \DecValTok{1} \DecValTok{1} \DecValTok{1} \DecValTok{1} \DecValTok{1} \DecValTok{1} \DecValTok{1}\NormalTok{ ...}
\NormalTok{\%}\SpecialCharTok{\textgreater{}}  \ErrorTok{$}\NormalTok{ IdLandkreis }\SpecialCharTok{:}\NormalTok{ int  }\DecValTok{1002} \DecValTok{1002} \DecValTok{1004} \DecValTok{1004} \DecValTok{1053} \DecValTok{1053} \DecValTok{1053} \DecValTok{1053} \DecValTok{1054} \DecValTok{1056}\NormalTok{ ...}
\NormalTok{\%}\SpecialCharTok{\textgreater{}}  \ErrorTok{$}\NormalTok{ time        }\SpecialCharTok{:}\NormalTok{ num  }\DecValTok{0} \DecValTok{0} \DecValTok{0} \DecValTok{0} \DecValTok{0} \DecValTok{0} \DecValTok{0} \DecValTok{0} \DecValTok{0} \DecValTok{0}\NormalTok{ ...}
\NormalTok{\%}\SpecialCharTok{\textgreater{}}  \ErrorTok{$}\NormalTok{ Age         }\SpecialCharTok{:}\NormalTok{ num  }\DecValTok{25} \DecValTok{47} \DecValTok{25} \DecValTok{47} \DecValTok{2} \DecValTok{25} \DecValTok{25} \DecValTok{70} \DecValTok{25} \DecValTok{10}\NormalTok{ ...}
\end{Highlighting}
\end{Shaded}

As illustrated by the output from above, we use the following simulation
data:

\begin{enumerate}
\item `Altersgruppe`: age group (intervals, categories), represented as character string
\item `Geschlecht`: gender 
\item `Day`: day of the infection
\item `IdBundesland`: federal state
\item `IdLandkreis`: county
\item `time`: number of days (`0` = start data). It will be used as `arrivalTimes` for the `simmer` simulations.
\item `Age`: integer representation of `Altersgruppe`
\end{enumerate}

\hypertarget{field-data}{%
\subsubsection{Field Data}\label{field-data}}

Next, we will describe the field data, i.e., the real ICU beds (or other
resources). Similar to the \texttt{rkidata}, which is available online
and can be downloaded from the RKI Server, the field data is also
available online. It can be downloaded from the DIVI Server as follows,
where \texttt{YYYY-MM-DD} should be replaced by the current date, e.g,
\texttt{2020-10-26}.

\begin{Shaded}
\begin{Highlighting}[]
\FunctionTok{updateIcudataFile}\NormalTok{(}\StringTok{"https://www.divi.de/joomlatools{-}files/docman{-}files/divi{-}intensivregister{-}tagesreports{-}csv/DIVI{-}Intensivregister\_YYYY{-}MM{-}DD\_12{-}15.csv"}\NormalTok{) }
\end{Highlighting}
\end{Shaded}

Note, the data structures on the DIVI server may change, so it might be
necessary to modify the following procedure. Please check the hints on
the DIVI web page. Contrary to the \texttt{updateRkidataFile()}
function, which downloads the complete historical dataset, the
\texttt{updateIcudataFile()} function only downloads data for a single
date. The downloaded data will be available in \texttt{babsim.hospital}
as \texttt{icudata}. As described in the Appendix (Section
\ref{sec:appendix}, the DIVI dataset is \emph{not} open data. Therefore,
only an example data set, that reflects the structure of the original
data from the DIVI register, is included in the \texttt{babsim.hospital}
package as \texttt{icudata}:

\begin{Shaded}
\begin{Highlighting}[]
\FunctionTok{str}\NormalTok{(babsim.hospital}\SpecialCharTok{::}\NormalTok{icudata)}
\NormalTok{\%}\SpecialCharTok{\textgreater{}} \StringTok{\textquotesingle{}data.frame\textquotesingle{}}\SpecialCharTok{:}    \DecValTok{40470}\NormalTok{ obs. of  }\DecValTok{9}\NormalTok{ variables}\SpecialCharTok{:}
\NormalTok{\%}\SpecialCharTok{\textgreater{}}  \ErrorTok{$}\NormalTok{ bundesland                  }\SpecialCharTok{:}\NormalTok{ int  }\DecValTok{1} \DecValTok{1} \DecValTok{1} \DecValTok{1} \DecValTok{1} \DecValTok{1} \DecValTok{1} \DecValTok{1} \DecValTok{1} \DecValTok{1}\NormalTok{ ...}
\NormalTok{\%}\SpecialCharTok{\textgreater{}}  \ErrorTok{$}\NormalTok{ gemeindeschluessel          }\SpecialCharTok{:}\NormalTok{ int  }\DecValTok{1001} \DecValTok{1002} \DecValTok{1003} \DecValTok{1004} \DecValTok{1051} \DecValTok{1053} \DecValTok{1054} \DecValTok{1055} \DecValTok{1056} \DecValTok{1057}\NormalTok{ ...}
\NormalTok{\%}\SpecialCharTok{\textgreater{}}  \ErrorTok{$}\NormalTok{ anzahl\_meldebereiche        }\SpecialCharTok{:}\NormalTok{ int  }\DecValTok{3} \DecValTok{5} \DecValTok{2} \DecValTok{1} \DecValTok{1} \DecValTok{2} \DecValTok{3} \DecValTok{3} \DecValTok{2} \DecValTok{1}\NormalTok{ ...}
\NormalTok{\%}\SpecialCharTok{\textgreater{}}  \ErrorTok{$}\NormalTok{ faelle\_covid\_aktuell        }\SpecialCharTok{:}\NormalTok{ int  }\DecValTok{0} \DecValTok{1} \DecValTok{0} \DecValTok{0} \DecValTok{0} \DecValTok{0} \DecValTok{0} \DecValTok{0} \DecValTok{1} \DecValTok{0}\NormalTok{ ...}
\NormalTok{\%}\SpecialCharTok{\textgreater{}}  \ErrorTok{$}\NormalTok{ faelle\_covid\_aktuell\_beatmet}\SpecialCharTok{:}\NormalTok{ int  }\DecValTok{0} \DecValTok{0} \DecValTok{0} \DecValTok{0} \DecValTok{0} \DecValTok{0} \DecValTok{0} \DecValTok{0} \DecValTok{0} \DecValTok{0}\NormalTok{ ...}
\NormalTok{\%}\SpecialCharTok{\textgreater{}}  \ErrorTok{$}\NormalTok{ anzahl\_standorte            }\SpecialCharTok{:}\NormalTok{ int  }\DecValTok{2} \DecValTok{3} \DecValTok{2} \DecValTok{1} \DecValTok{1} \DecValTok{2} \DecValTok{3} \DecValTok{3} \DecValTok{2} \DecValTok{1}\NormalTok{ ...}
\NormalTok{\%}\SpecialCharTok{\textgreater{}}  \ErrorTok{$}\NormalTok{ betten\_frei                 }\SpecialCharTok{:}\NormalTok{ int  }\DecValTok{24} \DecValTok{116} \DecValTok{103} \DecValTok{9} \DecValTok{13} \DecValTok{11} \DecValTok{15} \DecValTok{17} \DecValTok{9} \DecValTok{7}\NormalTok{ ...}
\NormalTok{\%}\SpecialCharTok{\textgreater{}}  \ErrorTok{$}\NormalTok{ betten\_belegt               }\SpecialCharTok{:}\NormalTok{ int  }\DecValTok{31} \DecValTok{113} \DecValTok{114} \DecValTok{16} \DecValTok{41} \DecValTok{13} \DecValTok{24} \DecValTok{35} \DecValTok{28} \DecValTok{5}\NormalTok{ ...}
\NormalTok{\%}\SpecialCharTok{\textgreater{}}  \ErrorTok{$}\NormalTok{ daten\_stand                 }\SpecialCharTok{:}\NormalTok{ Date, format}\SpecialCharTok{:} \StringTok{"2020{-}09{-}01"} \StringTok{"2020{-}09{-}01"}\NormalTok{ ...}
\end{Highlighting}
\end{Shaded}

The \texttt{ìcudata} can be visualized as follows (\texttt{region\ =\ 0}
is Germany, \texttt{region\ =\ 5} is North Rhine-Westphalia,
\texttt{region\ =\ 5374} is the Oberbergischer Kreis, etc.). Based on
the \texttt{icudata}, two plots can be generated. The first plot shows
the ICU beds without invasive ventilation. ICU beds without ventilation
can be calculated as
\texttt{faelle\_covid\_aktuell\ -\ faelle\_covid\_aktuell\_beatmet}.

\begin{Shaded}
\begin{Highlighting}[]
\NormalTok{p }\OtherTok{\textless{}{-}} \FunctionTok{ggVisualizeIcu}\NormalTok{(}\AttributeTok{region =} \DecValTok{5374}\NormalTok{)}
\FunctionTok{print}\NormalTok{(p[[}\DecValTok{1}\NormalTok{]])}
\end{Highlighting}
\end{Shaded}

\includegraphics[width=1\linewidth]{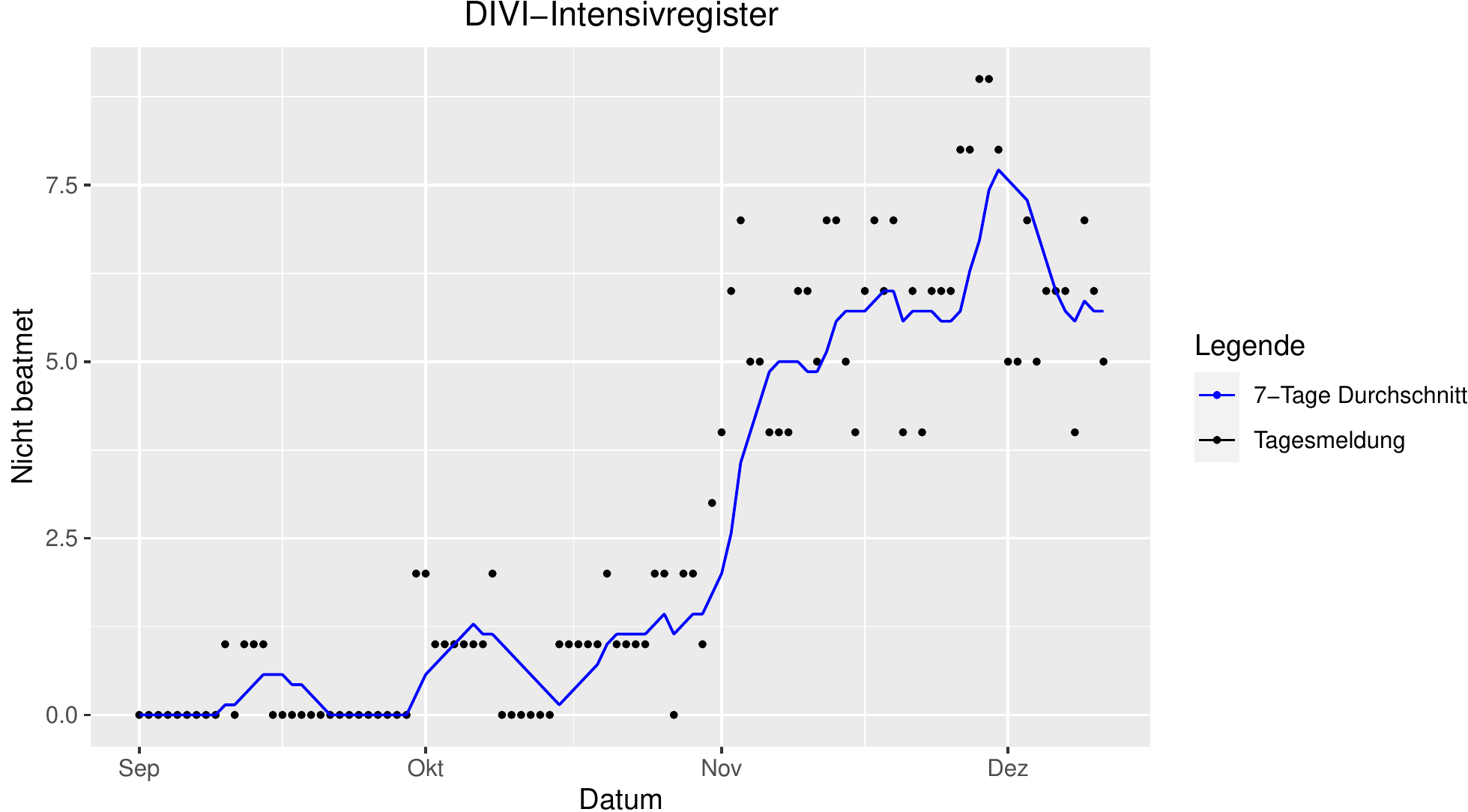}
The second plot shows ICU beds with ventilation:

\begin{Shaded}
\begin{Highlighting}[]
\FunctionTok{print}\NormalTok{(p[[}\DecValTok{2}\NormalTok{]])}
\end{Highlighting}
\end{Shaded}

\includegraphics[width=1\linewidth]{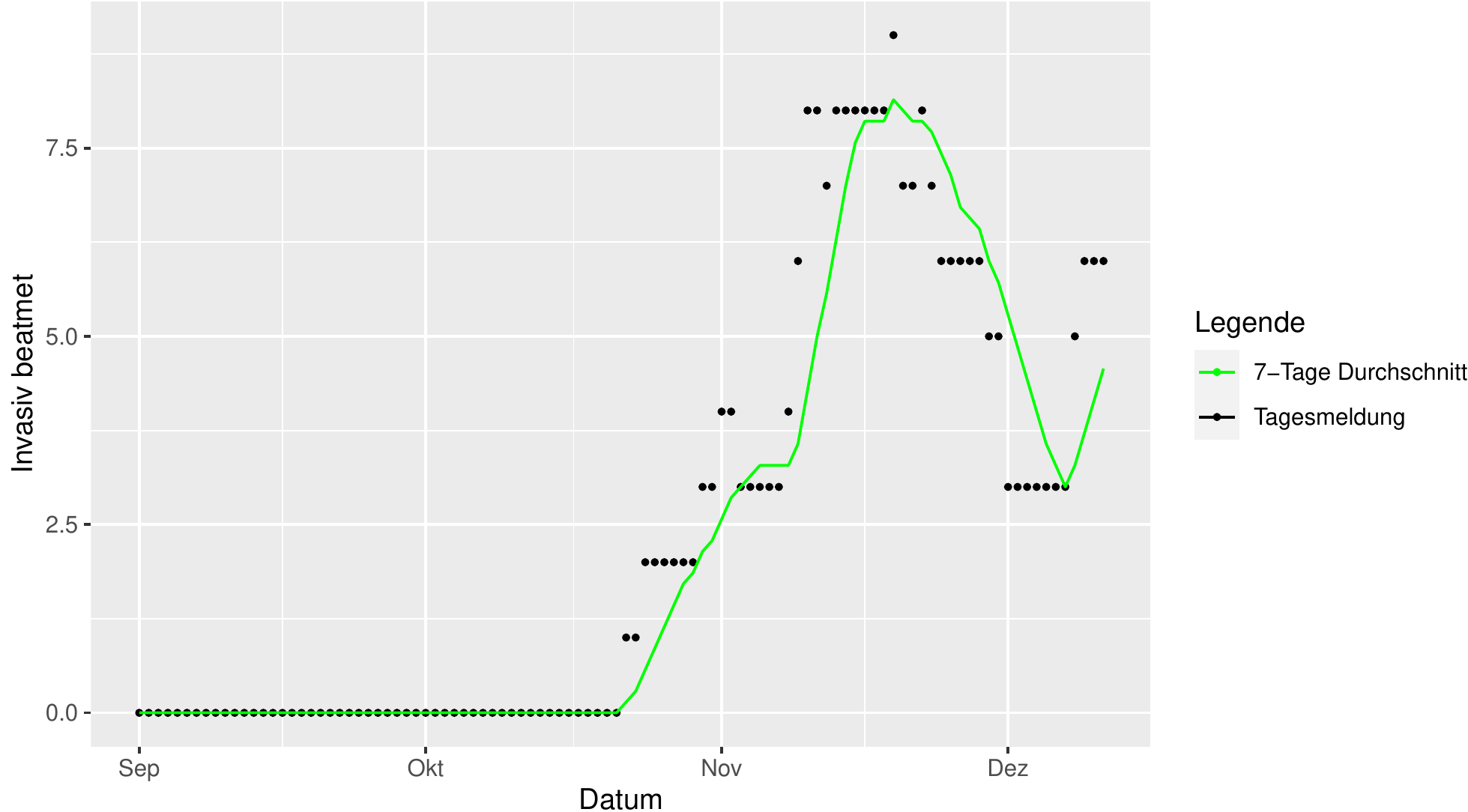}

The \texttt{icudata}, i.e., the field data or real data, will be
preprocessed as follows. The function \texttt{getIcuBeds()} converts the
9 dimensional DIVI ICU dataset \texttt{icudata}
(bundesland,gemeindeschluessel,\ldots, daten\_stand) into a data.frame
with two columns:

\begin{enumerate}
\item `intensiveBedVentilation`
\item `Day` 
\end{enumerate}

These are the two bed categories introduced above.

\begin{Shaded}
\begin{Highlighting}[]
\NormalTok{fieldData }\OtherTok{\textless{}{-}} \FunctionTok{getIcuBeds}\NormalTok{(babsim.hospital}\SpecialCharTok{::}\NormalTok{icudata)}
\FunctionTok{str}\NormalTok{(fieldData)}
\NormalTok{\%}\SpecialCharTok{\textgreater{}} \StringTok{\textquotesingle{}data.frame\textquotesingle{}}\SpecialCharTok{:}    \DecValTok{102}\NormalTok{ obs. of  }\DecValTok{3}\NormalTok{ variables}\SpecialCharTok{:}
\NormalTok{\%}\SpecialCharTok{\textgreater{}}  \ErrorTok{$}\NormalTok{ intensiveBed           }\SpecialCharTok{:}\NormalTok{ int  }\DecValTok{103} \DecValTok{103} \DecValTok{96} \DecValTok{97} \DecValTok{97} \DecValTok{92} \DecValTok{94} \DecValTok{100} \DecValTok{94} \DecValTok{104}\NormalTok{ ...}
\NormalTok{\%}\SpecialCharTok{\textgreater{}}  \ErrorTok{$}\NormalTok{ intensiveBedVentilation}\SpecialCharTok{:}\NormalTok{ int  }\DecValTok{132} \DecValTok{125} \DecValTok{127} \DecValTok{128} \DecValTok{126} \DecValTok{126} \DecValTok{134} \DecValTok{130} \DecValTok{133} \DecValTok{129}\NormalTok{ ...}
\NormalTok{\%}\SpecialCharTok{\textgreater{}}  \ErrorTok{$}\NormalTok{ Day                    }\SpecialCharTok{:}\NormalTok{ Date, format}\SpecialCharTok{:} \StringTok{"2020{-}09{-}01"} \StringTok{"2020{-}09{-}02"}\NormalTok{ ...}
\end{Highlighting}
\end{Shaded}

Now that we have introduced the required data, we can perform the first
simulations.

\hypertarget{performing-simulations}{%
\section{Performing Simulations}\label{performing-simulations}}

\label{sec:simulation}

To run a simulation, the setting must be configured (seed, number of
repeats, sequential or parallel evaluation, variable names, dates,
etc.). \texttt{babsim.hospital} uses a list to store information related
to the configuration. We will describe the components of this list
first. The configuration list contains information about the simulation
and field data.

\begin{Shaded}
\begin{Highlighting}[]
\NormalTok{region }\OtherTok{=} \DecValTok{5374}
\NormalTok{seed }\OtherTok{=} \DecValTok{123}
\NormalTok{simrepeats }\OtherTok{=} \DecValTok{2}
\NormalTok{parallel }\OtherTok{=} \ConstantTok{FALSE}
\NormalTok{percCores }\OtherTok{=} \FloatTok{0.8}
\NormalTok{resourceNames }\OtherTok{=}  \FunctionTok{c}\NormalTok{(}\StringTok{"intensiveBed"}\NormalTok{, }\StringTok{"intensiveBedVentilation"}\NormalTok{)}
\NormalTok{resourceEval }\OtherTok{=} \FunctionTok{c}\NormalTok{(}\StringTok{"intensiveBed"}\NormalTok{, }\StringTok{"intensiveBedVentilation"}\NormalTok{)}
\end{Highlighting}
\end{Shaded}

We can specify the field data based on \texttt{icudata} (DIVI) for the
simulation as follows:

\begin{Shaded}
\begin{Highlighting}[]
\NormalTok{FieldStartDate }\OtherTok{=} \StringTok{"2020{-}09{-}01"}
\NormalTok{icudata }\OtherTok{\textless{}{-}} \FunctionTok{getRegionIcu}\NormalTok{(}\AttributeTok{data =}\NormalTok{ icudata, }\AttributeTok{region =}\NormalTok{ region)}
\NormalTok{fieldData }\OtherTok{\textless{}{-}} \FunctionTok{getIcuBeds}\NormalTok{(icudata)}
\NormalTok{fieldData }\OtherTok{\textless{}{-}}\NormalTok{ fieldData[}\FunctionTok{which}\NormalTok{(fieldData}\SpecialCharTok{$}\NormalTok{Day }\SpecialCharTok{\textgreater{}=} \FunctionTok{as.Date}\NormalTok{(FieldStartDate)), ]}
\FunctionTok{rownames}\NormalTok{(fieldData) }\OtherTok{\textless{}{-}} \ConstantTok{NULL}
\NormalTok{icu }\OtherTok{=} \ConstantTok{TRUE}
\NormalTok{icuWeights }\OtherTok{=} \FunctionTok{c}\NormalTok{(}\DecValTok{1}\NormalTok{,}\DecValTok{1}\NormalTok{)}
\end{Highlighting}
\end{Shaded}

Next, simulation data (RKI data) can be selected. The simulation data in
our example, depend on the field data:

\begin{Shaded}
\begin{Highlighting}[]
\NormalTok{SimStartDate }\OtherTok{=} \StringTok{"2020{-}08{-}01"}
\NormalTok{rkidata }\OtherTok{\textless{}{-}} \FunctionTok{getRegionRki}\NormalTok{(}\AttributeTok{data =}\NormalTok{ rkidata, }\AttributeTok{region =}\NormalTok{ region)}
\NormalTok{simData }\OtherTok{\textless{}{-}} \FunctionTok{getRkiData}\NormalTok{(rkidata)}
\NormalTok{simData }\OtherTok{\textless{}{-}}\NormalTok{ simData[}\FunctionTok{which}\NormalTok{(simData}\SpecialCharTok{$}\NormalTok{Day }\SpecialCharTok{\textgreater{}=} \FunctionTok{as.Date}\NormalTok{(SimStartDate)), ]}
\NormalTok{simData }\OtherTok{\textless{}{-}}\NormalTok{ simData[}\FunctionTok{as.Date}\NormalTok{(simData}\SpecialCharTok{$}\NormalTok{Day) }\SpecialCharTok{\textless{}=} \FunctionTok{max}\NormalTok{(}\FunctionTok{as.Date}\NormalTok{(fieldData}\SpecialCharTok{$}\NormalTok{Day)),]}
\NormalTok{simData}\SpecialCharTok{$}\NormalTok{time }\OtherTok{\textless{}{-}}\NormalTok{ simData}\SpecialCharTok{$}\NormalTok{time }\SpecialCharTok{{-}} \FunctionTok{min}\NormalTok{(simData}\SpecialCharTok{$}\NormalTok{time)}
\FunctionTok{rownames}\NormalTok{(simData) }\OtherTok{\textless{}{-}} \ConstantTok{NULL}
\end{Highlighting}
\end{Shaded}

Finally, we combine all field and simulation data into a single
\texttt{list()} called \texttt{data}:

\begin{Shaded}
\begin{Highlighting}[]
\NormalTok{data }\OtherTok{\textless{}{-}} \FunctionTok{list}\NormalTok{(}\AttributeTok{simData =}\NormalTok{ simData, }\AttributeTok{fieldData =}\NormalTok{ fieldData)}
\end{Highlighting}
\end{Shaded}

Configuration information is stored in the \texttt{conf} list, i.e.,
\texttt{conf} refers to the simulation configuration, e.g., sequential
or parallel evaluation, number of cores, resource names, log level, etc.

\begin{Shaded}
\begin{Highlighting}[]
\NormalTok{conf }\OtherTok{\textless{}{-}} \FunctionTok{babsimToolsConf}\NormalTok{()}
\NormalTok{conf }\OtherTok{\textless{}{-}} \FunctionTok{getConfFromData}\NormalTok{(}\AttributeTok{conf =}\NormalTok{ conf,}
                        \AttributeTok{simData =}\NormalTok{ data}\SpecialCharTok{$}\NormalTok{simData,}
                        \AttributeTok{fieldData =}\NormalTok{ data}\SpecialCharTok{$}\NormalTok{fieldData)}
\NormalTok{conf}\SpecialCharTok{$}\NormalTok{parallel }\OtherTok{=}\NormalTok{ parallel}
\NormalTok{conf}\SpecialCharTok{$}\NormalTok{simRepeats }\OtherTok{=}\NormalTok{ simrepeats}
\NormalTok{conf}\SpecialCharTok{$}\NormalTok{ICU }\OtherTok{=}\NormalTok{ icu}
\NormalTok{conf}\SpecialCharTok{$}\NormalTok{ResourceNames }\OtherTok{=}\NormalTok{ resourceNames}
\NormalTok{conf}\SpecialCharTok{$}\NormalTok{ResourceEval }\OtherTok{=}\NormalTok{ resourceEval}
\NormalTok{conf}\SpecialCharTok{$}\NormalTok{percCores }\OtherTok{=}\NormalTok{ percCores}
\NormalTok{conf}\SpecialCharTok{$}\NormalTok{logLevel }\OtherTok{=} \DecValTok{1}
\NormalTok{conf}\SpecialCharTok{$}\NormalTok{w2 }\OtherTok{=}\NormalTok{ icuWeights}
\FunctionTok{set.seed}\NormalTok{(conf}\SpecialCharTok{$}\NormalTok{seed)}
\end{Highlighting}
\end{Shaded}

In addition to the configuration list, a second list, which stores
information about the simulation model parameters, is used. The core of
the \texttt{babsim.hospital} simulations is based on the
\href{https://r-simmer.org/}{\texttt{simmer}} package. It uses
simulation parameters, e.g., arrival times, durations, and transition
probabilities. There are currently 29 parameters (shown below) that are
stored in the list \texttt{para}.

\begin{Shaded}
\begin{Highlighting}[]
\NormalTok{para }\OtherTok{\textless{}{-}} \FunctionTok{babsimHospitalPara}\NormalTok{()}
\FunctionTok{str}\NormalTok{(para)}
\NormalTok{\%}\SpecialCharTok{\textgreater{}}\NormalTok{ List of }\DecValTok{29}
\NormalTok{\%}\SpecialCharTok{\textgreater{}}  \ErrorTok{$}\NormalTok{ AmntDaysInfectedToHospital               }\SpecialCharTok{:}\NormalTok{ num }\FloatTok{9.5}
\NormalTok{\%}\SpecialCharTok{\textgreater{}}  \ErrorTok{$}\NormalTok{ AmntDaysNormalToHealthy                  }\SpecialCharTok{:}\NormalTok{ num }\DecValTok{10}
\NormalTok{\%}\SpecialCharTok{\textgreater{}}  \ErrorTok{$}\NormalTok{ AmntDaysNormalToIntensive                }\SpecialCharTok{:}\NormalTok{ num }\DecValTok{5}
\NormalTok{\%}\SpecialCharTok{\textgreater{}}  \ErrorTok{$}\NormalTok{ AmntDaysNormalToVentilation              }\SpecialCharTok{:}\NormalTok{ num }\FloatTok{3.63}
\NormalTok{\%}\SpecialCharTok{\textgreater{}}  \ErrorTok{$}\NormalTok{ AmntDaysNormalToDeath                    }\SpecialCharTok{:}\NormalTok{ num }\DecValTok{5}
\NormalTok{\%}\SpecialCharTok{\textgreater{}}  \ErrorTok{$}\NormalTok{ AmntDaysIntensiveToAftercare             }\SpecialCharTok{:}\NormalTok{ num }\DecValTok{7}
\NormalTok{\%}\SpecialCharTok{\textgreater{}}  \ErrorTok{$}\NormalTok{ AmntDaysIntensiveToVentilation           }\SpecialCharTok{:}\NormalTok{ num }\DecValTok{4}
\NormalTok{\%}\SpecialCharTok{\textgreater{}}  \ErrorTok{$}\NormalTok{ AmntDaysIntensiveToDeath                 }\SpecialCharTok{:}\NormalTok{ num }\DecValTok{5}
\NormalTok{\%}\SpecialCharTok{\textgreater{}}  \ErrorTok{$}\NormalTok{ AmntDaysVentilationToIntensiveAfter      }\SpecialCharTok{:}\NormalTok{ num }\DecValTok{30}
\NormalTok{\%}\SpecialCharTok{\textgreater{}}  \ErrorTok{$}\NormalTok{ AmntDaysVentilationToDeath               }\SpecialCharTok{:}\NormalTok{ num }\DecValTok{20}
\NormalTok{\%}\SpecialCharTok{\textgreater{}}  \ErrorTok{$}\NormalTok{ AmntDaysIntensiveAfterToAftercare        }\SpecialCharTok{:}\NormalTok{ num }\DecValTok{3}
\NormalTok{\%}\SpecialCharTok{\textgreater{}}  \ErrorTok{$}\NormalTok{ AmntDaysIntensiveAfterToDeath            }\SpecialCharTok{:}\NormalTok{ num }\DecValTok{4}
\NormalTok{\%}\SpecialCharTok{\textgreater{}}  \ErrorTok{$}\NormalTok{ GammaShapeParameter                      }\SpecialCharTok{:}\NormalTok{ num }\DecValTok{1}
\NormalTok{\%}\SpecialCharTok{\textgreater{}}  \ErrorTok{$}\NormalTok{ FactorPatientsInfectedToHospital         }\SpecialCharTok{:}\NormalTok{ num }\FloatTok{0.1}
\NormalTok{\%}\SpecialCharTok{\textgreater{}}  \ErrorTok{$}\NormalTok{ FactorPatientsHospitalToIntensive        }\SpecialCharTok{:}\NormalTok{ num }\FloatTok{0.09}
\NormalTok{\%}\SpecialCharTok{\textgreater{}}  \ErrorTok{$}\NormalTok{ FactorPatientsHospitalToVentilation      }\SpecialCharTok{:}\NormalTok{ num }\FloatTok{0.01}
\NormalTok{\%}\SpecialCharTok{\textgreater{}}  \ErrorTok{$}\NormalTok{ FactorPatientsNormalToIntensive          }\SpecialCharTok{:}\NormalTok{ num }\FloatTok{0.1}
\NormalTok{\%}\SpecialCharTok{\textgreater{}}  \ErrorTok{$}\NormalTok{ FactorPatientsNormalToVentilation        }\SpecialCharTok{:}\NormalTok{ num }\FloatTok{0.001}
\NormalTok{\%}\SpecialCharTok{\textgreater{}}  \ErrorTok{$}\NormalTok{ FactorPatientsNormalToDeath              }\SpecialCharTok{:}\NormalTok{ num }\FloatTok{0.1}
\NormalTok{\%}\SpecialCharTok{\textgreater{}}  \ErrorTok{$}\NormalTok{ FactorPatientsIntensiveToVentilation     }\SpecialCharTok{:}\NormalTok{ num }\FloatTok{0.3}
\NormalTok{\%}\SpecialCharTok{\textgreater{}}  \ErrorTok{$}\NormalTok{ FactorPatientsIntensiveToDeath           }\SpecialCharTok{:}\NormalTok{ num }\FloatTok{0.1}
\NormalTok{\%}\SpecialCharTok{\textgreater{}}  \ErrorTok{$}\NormalTok{ FactorPatientsVentilationToIntensiveAfter}\SpecialCharTok{:}\NormalTok{ num }\FloatTok{0.7}
\NormalTok{\%}\SpecialCharTok{\textgreater{}}  \ErrorTok{$}\NormalTok{ FactorPatientsIntensiveAfterToDeath      }\SpecialCharTok{:}\NormalTok{ num }\FloatTok{1e{-}05}
\NormalTok{\%}\SpecialCharTok{\textgreater{}}  \ErrorTok{$}\NormalTok{ AmntDaysAftercareToHealthy               }\SpecialCharTok{:}\NormalTok{ num }\DecValTok{3}
\NormalTok{\%}\SpecialCharTok{\textgreater{}}  \ErrorTok{$}\NormalTok{ RiskFactorA                              }\SpecialCharTok{:}\NormalTok{ num }\FloatTok{0.0205}
\NormalTok{\%}\SpecialCharTok{\textgreater{}}  \ErrorTok{$}\NormalTok{ RiskFactorB                              }\SpecialCharTok{:}\NormalTok{ num }\FloatTok{0.01}
\NormalTok{\%}\SpecialCharTok{\textgreater{}}  \ErrorTok{$}\NormalTok{ RiskMale                                 }\SpecialCharTok{:}\NormalTok{ num }\FloatTok{1.5}
\NormalTok{\%}\SpecialCharTok{\textgreater{}}  \ErrorTok{$}\NormalTok{ AmntDaysIntensiveAfterToHealthy          }\SpecialCharTok{:}\NormalTok{ num }\DecValTok{3}
\NormalTok{\%}\SpecialCharTok{\textgreater{}}  \ErrorTok{$}\NormalTok{ FactorPatientsIntensiveAfterToHealthy    }\SpecialCharTok{:}\NormalTok{ num }\FloatTok{0.67}
\end{Highlighting}
\end{Shaded}

\hypertarget{simulation-runs}{%
\subsection{Simulation Runs}\label{simulation-runs}}

The \texttt{babsim.hospital} simulator requires the specification of

\begin{enumerate}
\item  `arrivalTimes`
\item configuration list `conf`
\item  parameter list `para` 
\end{enumerate}

for the simulation. Arrival times were not discussed yet.
\texttt{babsim.hospital} provides the function \texttt{getRkiRisk()}
that generates arrivals with associated risks. The \texttt{Risk} is
based on age (\texttt{Altersgruppe}) and gender (\texttt{Geschlect}):

\begin{Shaded}
\begin{Highlighting}[]
\NormalTok{rkiWithRisk }\OtherTok{\textless{}{-}} \FunctionTok{getRkiRisk}\NormalTok{(data}\SpecialCharTok{$}\NormalTok{simData, para)}
\FunctionTok{head}\NormalTok{(rkiWithRisk)}
\NormalTok{\%}\SpecialCharTok{\textgreater{}}\NormalTok{   Altersgruppe Geschlecht        Day IdBundesland IdLandkreis time Age}
\NormalTok{\%}\SpecialCharTok{\textgreater{}} \DecValTok{1}\NormalTok{      A15}\SpecialCharTok{{-}}\NormalTok{A34          M }\DecValTok{2020{-}09{-}01}            \DecValTok{5}        \DecValTok{5374}    \DecValTok{0}  \DecValTok{25}
\NormalTok{\%}\SpecialCharTok{\textgreater{}} \DecValTok{2}\NormalTok{      A15}\SpecialCharTok{{-}}\NormalTok{A34          M }\DecValTok{2020{-}09{-}01}            \DecValTok{5}        \DecValTok{5374}    \DecValTok{0}  \DecValTok{25}
\NormalTok{\%}\SpecialCharTok{\textgreater{}} \DecValTok{3}\NormalTok{      A15}\SpecialCharTok{{-}}\NormalTok{A34          M }\DecValTok{2020{-}09{-}01}            \DecValTok{5}        \DecValTok{5374}    \DecValTok{0}  \DecValTok{25}
\NormalTok{\%}\SpecialCharTok{\textgreater{}} \DecValTok{4}\NormalTok{      A35}\SpecialCharTok{{-}}\NormalTok{A59          M }\DecValTok{2020{-}09{-}01}            \DecValTok{5}        \DecValTok{5374}    \DecValTok{0}  \DecValTok{47}
\NormalTok{\%}\SpecialCharTok{\textgreater{}} \DecValTok{5}\NormalTok{      A35}\SpecialCharTok{{-}}\NormalTok{A59          W }\DecValTok{2020{-}09{-}01}            \DecValTok{5}        \DecValTok{5374}    \DecValTok{0}  \DecValTok{47}
\NormalTok{\%}\SpecialCharTok{\textgreater{}} \DecValTok{6}\NormalTok{      A35}\SpecialCharTok{{-}}\NormalTok{A59          W }\DecValTok{2020{-}09{-}01}            \DecValTok{5}        \DecValTok{5374}    \DecValTok{0}  \DecValTok{47}
\NormalTok{\%}\SpecialCharTok{\textgreater{}}\NormalTok{         Risk}
\NormalTok{\%}\SpecialCharTok{\textgreater{}} \DecValTok{1} \FloatTok{0.03946352}
\NormalTok{\%}\SpecialCharTok{\textgreater{}} \DecValTok{2} \FloatTok{0.03946352}
\NormalTok{\%}\SpecialCharTok{\textgreater{}} \DecValTok{3} \FloatTok{0.03946352}
\NormalTok{\%}\SpecialCharTok{\textgreater{}} \DecValTok{4} \FloatTok{0.04917457}
\NormalTok{\%}\SpecialCharTok{\textgreater{}} \DecValTok{5} \FloatTok{0.03278305}
\NormalTok{\%}\SpecialCharTok{\textgreater{}} \DecValTok{6} \FloatTok{0.03278305}
\end{Highlighting}
\end{Shaded}

To perform simulations, only two parameters are required:

\begin{enumerate}
\item  `time`: arrival time 
\item  `Risk`: risk (based on age and gender)
\end{enumerate}

A data.frame with these two parameters is passed to the main simulation
function \texttt{babsimHospital}. Output from the simulation is stored
in the variable \texttt{envs}. The simulation run is started as follows:

\hypertarget{visualize-and-evaluate-simulation-output}{%
\section{Visualize and Evaluate Simulation
Output}\label{visualize-and-evaluate-simulation-output}}

\label{sec:visualization}

\hypertarget{simmer-plots}{%
\subsection{Simmer Plots}\label{simmer-plots}}

First, we illustrate how to generate plots using the
\texttt{simmer.plot} package (Ucar, Smeets, and Azcorra 2019). In the
following graph, the individual lines are all separate replications. The
smoothing performed is a cumulative average. Besides
\texttt{intensiveBed} and \texttt{intensiveBedVentilation},
\texttt{babsim.hospital} also provides information about the number of
non-ICU beds. The non-ICU beds are labeled as \texttt{bed}. Summarizing,
\texttt{babsim.hospital} generates output for three bed categories:

\begin{enumerate}
\item `bed`
\item `intensiveBed`
\item `intensiveBedVentilation`
\end{enumerate}

To plot resource usage for three resources side-by-side, we can proceed
as follows:

\begin{Shaded}
\begin{Highlighting}[]
\NormalTok{resources }\OtherTok{\textless{}{-}} \FunctionTok{get\_mon\_resources}\NormalTok{(envs)}
\NormalTok{resources}\SpecialCharTok{$}\NormalTok{capacity }\OtherTok{\textless{}{-}}\NormalTok{ resources}\SpecialCharTok{$}\NormalTok{capacity}\SpecialCharTok{/}\FloatTok{1e5}
\FunctionTok{plot}\NormalTok{(resources, }\AttributeTok{metric =} \StringTok{"usage"}\NormalTok{, }\FunctionTok{c}\NormalTok{(}\StringTok{"bed"}\NormalTok{, }\StringTok{"intensiveBed"}\NormalTok{, }\StringTok{"intensiveBedVentilation"}\NormalTok{), }\AttributeTok{items =} \StringTok{"server"}\NormalTok{)}
\end{Highlighting}
\end{Shaded}

\includegraphics{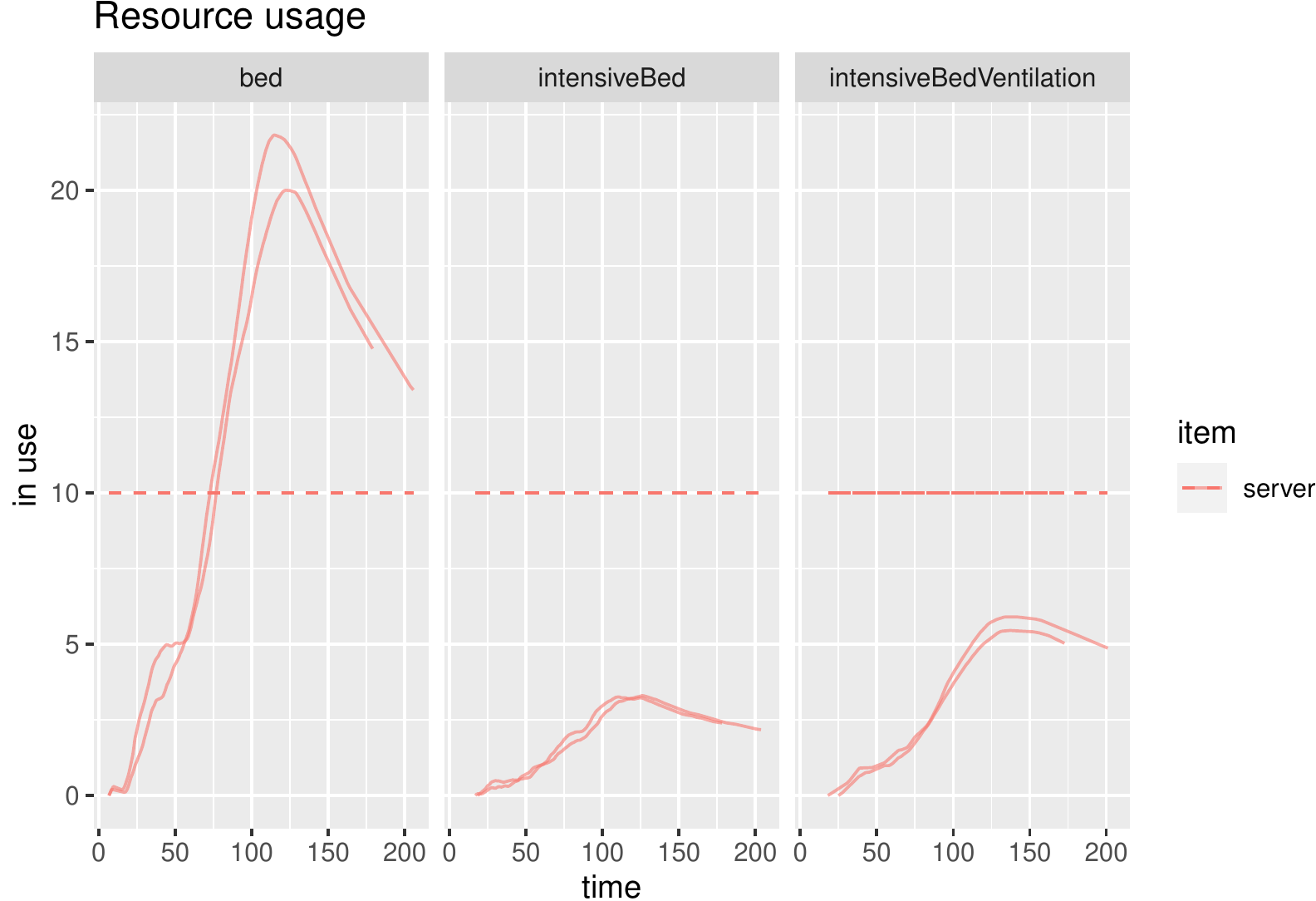}
Note, each resource can be plotted separately. For example, the
following command generates a plot of non icu beds.

\begin{Shaded}
\begin{Highlighting}[]
\FunctionTok{plot}\NormalTok{(resources, }\AttributeTok{metric =} \StringTok{"usage"}\NormalTok{, }\StringTok{"bed"}\NormalTok{, }\AttributeTok{items =} \StringTok{"server"}\NormalTok{, }\AttributeTok{steps =} \ConstantTok{TRUE}\NormalTok{)}
\end{Highlighting}
\end{Shaded}

\includegraphics{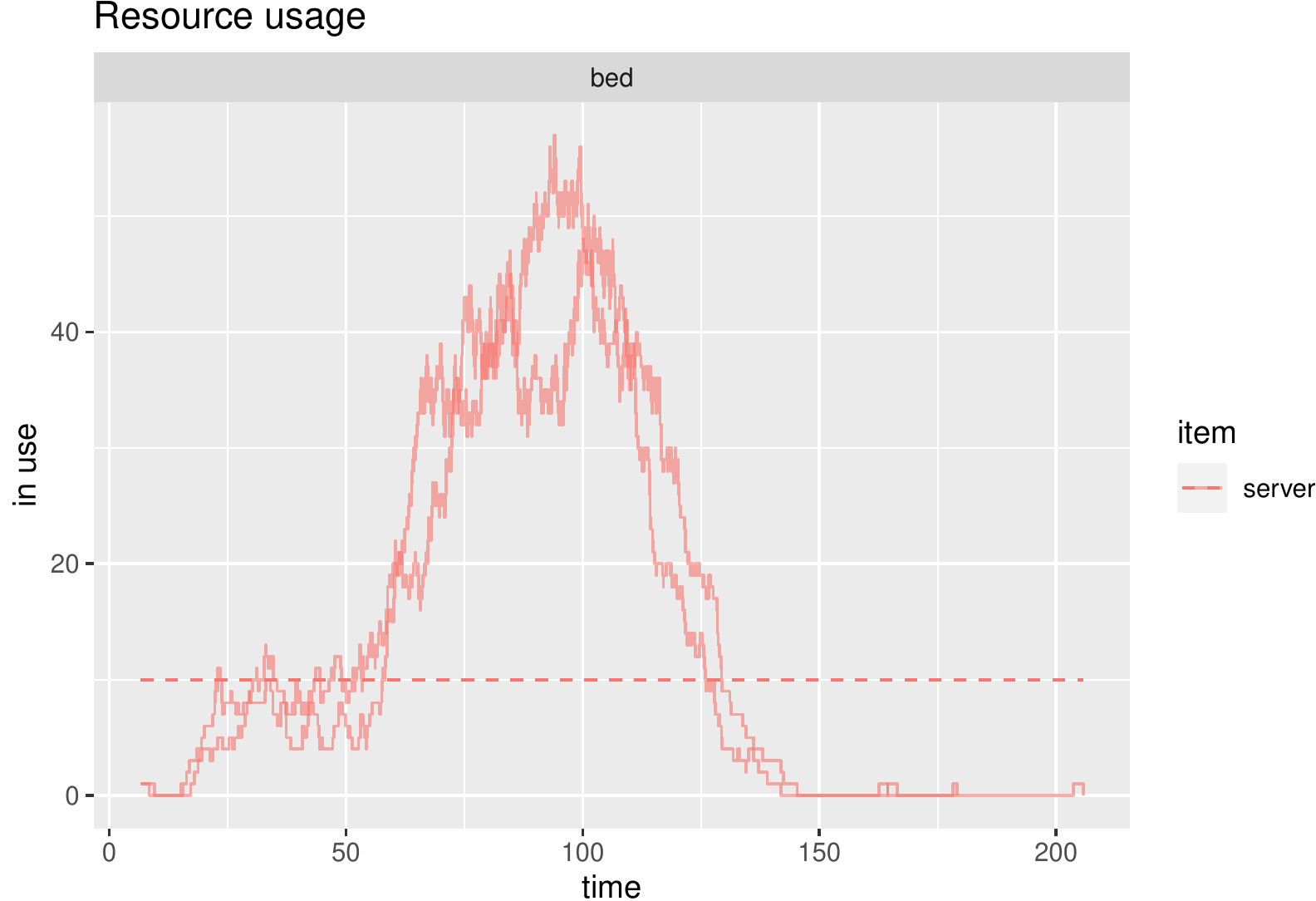}

\hypertarget{evaluation-of-simulation-results}{%
\subsection{Evaluation of Simulation
Results}\label{evaluation-of-simulation-results}}

\texttt{babsim.hospital} provides functions for evaluating the quality
of the simulation results. Simulation results depend on the transition
probabilities and durations, i.e., a vector of more than 30 variables.
These vectors represent \emph{parameter settings}.
\texttt{babsim.hospital} provides a \emph{default} parameter set, that
is based on knowledge from domain experts (doctors, members of COVID-19
crises teams, mathematicians, and many more). We can calculate the error
(RMSE) of the default parameter setting, which was used in this
simulation, as follows:

\begin{Shaded}
\begin{Highlighting}[]
\NormalTok{fieldEvents }\OtherTok{\textless{}{-}} \FunctionTok{getRealBeds}\NormalTok{(}\AttributeTok{data =}\NormalTok{ data}\SpecialCharTok{$}\NormalTok{fieldData,}
                        \AttributeTok{resource =}\NormalTok{ conf}\SpecialCharTok{$}\NormalTok{ResourceNames)}
\NormalTok{res }\OtherTok{\textless{}{-}} \FunctionTok{getDailyMaxResults}\NormalTok{(}\AttributeTok{envs =}\NormalTok{ envs,  }\AttributeTok{fieldEvents =}\NormalTok{ fieldEvents, }\AttributeTok{conf=}\NormalTok{conf)}
\NormalTok{resDefault }\OtherTok{\textless{}{-}} \FunctionTok{getError}\NormalTok{(res, }\AttributeTok{conf=}\NormalTok{conf)}
\FunctionTok{print}\NormalTok{(resDefault)}
\NormalTok{\%}\SpecialCharTok{\textgreater{}}\NormalTok{ [}\DecValTok{1}\NormalTok{] }\FloatTok{9.412271}
\end{Highlighting}
\end{Shaded}

The error is 9.4122709.

\hypertarget{generating-babsim.hospital-plots}{%
\subsection{Generating babsim.hospital
Plots}\label{generating-babsim.hospital-plots}}

In addition to the original \texttt{simmer} plot,
\texttt{babsim.hospital} provides functions for visualization. Here, we
illustrate how \texttt{babsim.hospital} plots can be generated. Before
we deescribe these plots, readers should be aware of the fact, that we
do not use the full data, simulation results are completely wrong and do
not represent any real-world situation! The following figures are
included to demonstrate the working principles of the visualization
procedures.

\begin{Shaded}
\begin{Highlighting}[]
\NormalTok{p }\OtherTok{\textless{}{-}} \FunctionTok{plotDailyMaxResults}\NormalTok{(res)}
\FunctionTok{plot}\NormalTok{(p)}
\end{Highlighting}
\end{Shaded}

\includegraphics{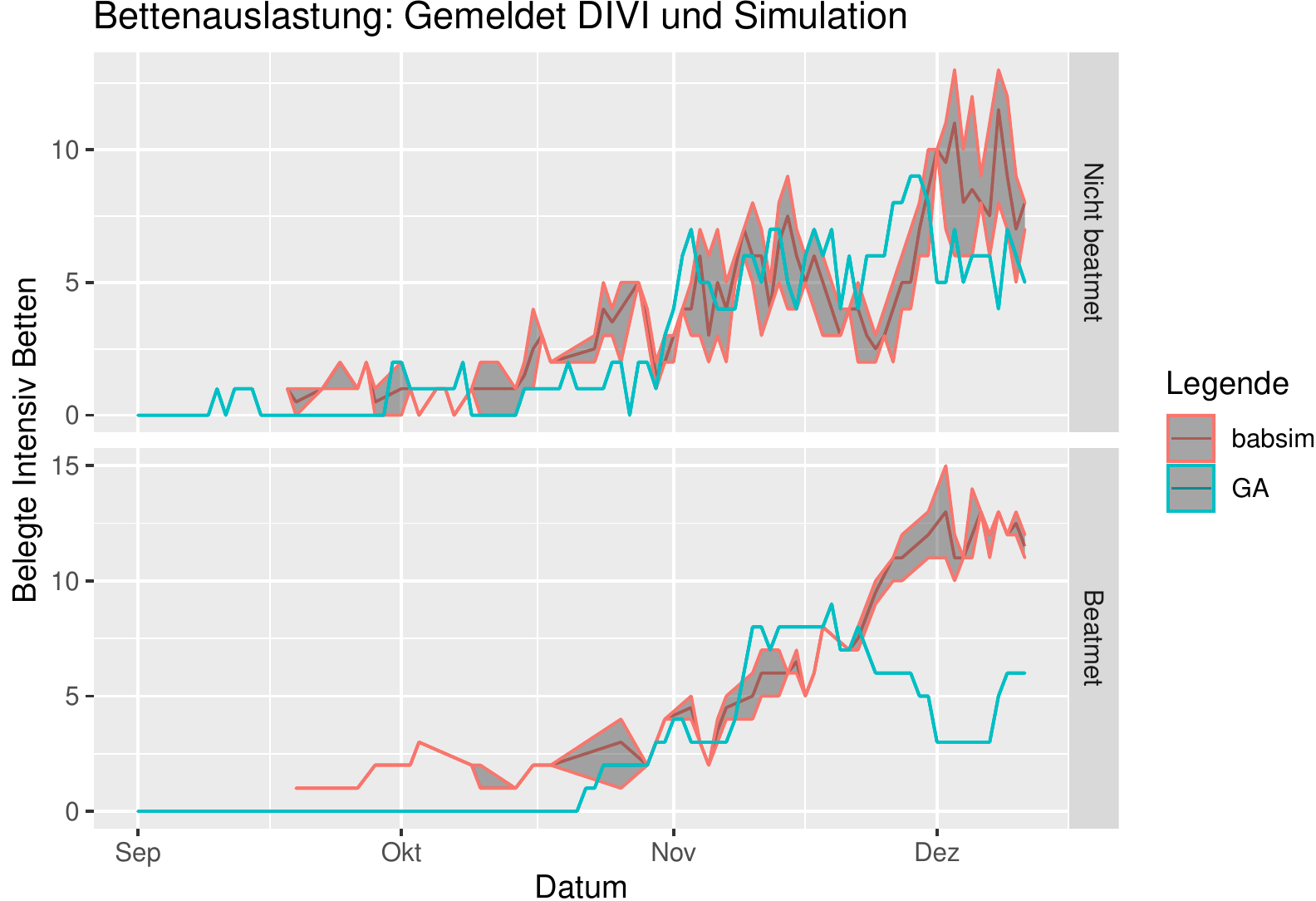}

Note, \texttt{ggplot} and \texttt{plotly}can be used as follows to
generate interactive plots.

\begin{Shaded}
\begin{Highlighting}[]
\FunctionTok{ggplotly}\NormalTok{(p) }
\end{Highlighting}
\end{Shaded}

\hypertarget{optimization}{%
\section{Optimization}\label{optimization}}

\label{sec:optimization} As discussed above, \texttt{babsim.hospital}
provides a default parameter set, which can be used for simulations. The
function \texttt{babsimHospitalPara()} provides a convenient way to
access the default parameter set:

\begin{Shaded}
\begin{Highlighting}[]
\NormalTok{para }\OtherTok{\textless{}{-}} \FunctionTok{babsimHospitalPara}\NormalTok{()}
\end{Highlighting}
\end{Shaded}

\texttt{babsim.hospital} provides an interface to optimize the parameter
values of the simulation model. The following code is just a quick demo.
To run the following code, the complete \texttt{rkidata} and
\texttt{icudata} data sets must be available. Please download the data
from RKI and DIVI or provide your own simulation and field data! Note:
results are stored in the directory \texttt{results}.

\begin{verbatim}
%> [1] "2020/09/01 00:00:00" "2020/12/11 00:00:00"
%> [1] "2020-09-01" "2020-12-11"
%> [1] "trainDataSim:  2020-10-05" "trainDataSim:  2020-12-11"
%> [1] "trainDataField:  2020-11-02" "trainDataField:  2020-12-11"
%> [1] "Starting optimization loop:"
%> [1] "#########################################"
%> [1] "Repeat: 1 ###############################"
%> [1] "trainConfSim:  2020-10-05" "trainConfSim:  2020-12-11"
%> [1] "trainConfField:  2020-11-02" "trainConfField:  2020-12-11"
%> [1] "Warning cutting some x0 parameters as there are too many"
%> [1] "Starting Surrogate Optimization"
%> [1] "*******************************"
%> 60% completed.
%> [1] 79
%> [1] "results/test_2020_Dez.14_00.54_V11.5.16R1.RData"
%> [1] "results/test_2020_Dez.14_00.54_V11.5.16R.RData"
\end{verbatim}

The code from above was shown for didactical purposes. It starts a short
optimization run to illustrate the underlying optimization procedure.
Results from real the \texttt{runoptDirect()} runs are stored in the
\texttt{paras.rda} file. \texttt{babsim.hospital} provides results from
several regions (towns and counties in Germany), e.g.:

\begin{itemize}
\item `getParaSet(5374)`: Oberbergischer Kreis
\item `getParaSet(5315)`: City of Cologne
\item `getParaSet(5)`: North-Rhine Westphalia
\item `getParaSet(0)`: Germany
\end{itemize}

\hypertarget{use-optimized-parameters}{%
\subsection{Use Optimized Parameters}\label{use-optimized-parameters}}

Results, i.e., parameter settings, of the short \texttt{runoptDirect()}
optimization from above can be used as follows:

\begin{Shaded}
\begin{Highlighting}[]
\NormalTok{xy }\OtherTok{\textless{}{-}}\NormalTok{ resDemo}\SpecialCharTok{$}\NormalTok{best.df}
\NormalTok{para }\OtherTok{\textless{}{-}} \FunctionTok{getBestParameter}\NormalTok{(xy)}
\NormalTok{res }\OtherTok{\textless{}{-}} \FunctionTok{modelResultHospital}\NormalTok{(}\AttributeTok{para=}\NormalTok{para, }
                           \AttributeTok{conf=}\NormalTok{conf,}
                           \AttributeTok{data =}\NormalTok{ data)}

\NormalTok{resOpt }\OtherTok{\textless{}{-}} \FunctionTok{getError}\NormalTok{(res, }\AttributeTok{conf=}\NormalTok{conf)}
\FunctionTok{print}\NormalTok{(resOpt)}
\NormalTok{\%}\SpecialCharTok{\textgreater{}}\NormalTok{ [}\DecValTok{1}\NormalTok{] }\FloatTok{6.566368}
\end{Highlighting}
\end{Shaded}

These results show that even a very short optimization improves the
error: The improved is 6.5663677, which is smaller than 9.4122709. This
improvement can also be visualized.

\begin{Shaded}
\begin{Highlighting}[]
\NormalTok{p }\OtherTok{\textless{}{-}} \FunctionTok{plotDailyMaxResults}\NormalTok{(res)}
\FunctionTok{print}\NormalTok{(p)}
\end{Highlighting}
\end{Shaded}

\includegraphics{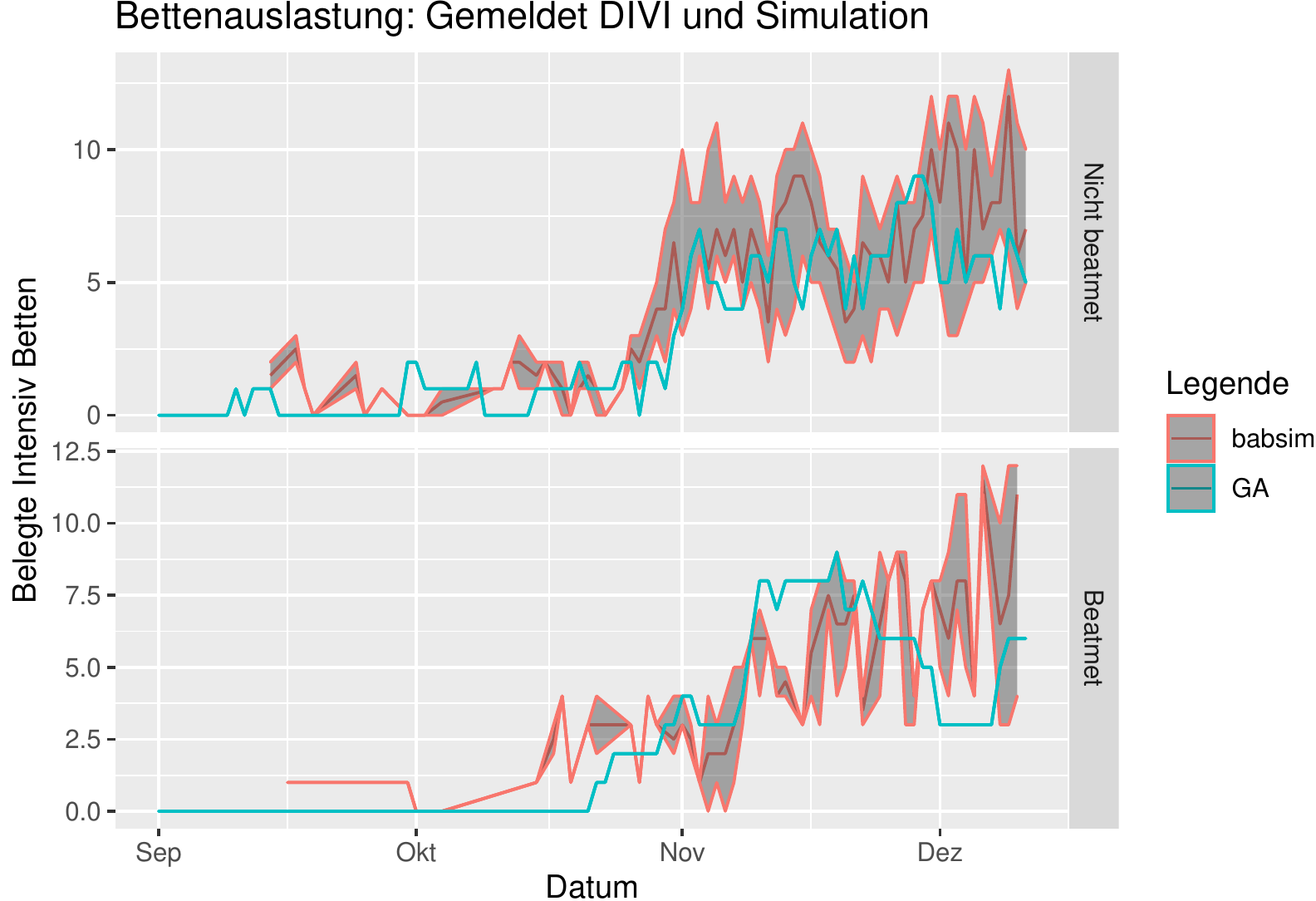}

\hypertarget{visualize-parameter-settings}{%
\section{Visualize Parameter
Settings}\label{visualize-parameter-settings}}

\label{sec:vispara} Besides using the optimized parameters for
simulations, which allows improved simulations, the optimized parameters
can be used for analysing the parameter settings.
\texttt{babsim.hospital} includes several tools to analyze parameter
settings. You might recall that parameter settings consist of

\begin{itemize}
\item transition probabilities, e.g., the probability that an infected
   individual has to go to the hospital. 
\item durations, e.g., the time span until an infected individual goes to the
   hospital (in days).
\end{itemize}

The following plot illustrates the transition probabilities. It uses the
following states:

\begin{enumerate}
\item `infec`: infected
\item `out`: transfer out, no hospital required
\item `hosp`: hospital
\item `normal`: normal station, no ICU
\item `intens`: ICU (without ventilation)
\item `vent`: ICU ventilated
\item `intafter`: intensive aftercare (from ICU with ventilation, on ICU)
\item `aftercare`: aftercare (from ICU,  on normal station)
\item `death`: patient dies
\item `healthy`: recovered
\end{enumerate}

The transition matrix, that stores the probabilities, is shown below:

\begin{Shaded}
\begin{Highlighting}[]
\NormalTok{para }\OtherTok{\textless{}{-}} \FunctionTok{babsimHospitalPara}\NormalTok{()}
\FunctionTok{getMatrixP}\NormalTok{(}\AttributeTok{para =}\NormalTok{ para)}
\NormalTok{\%}\SpecialCharTok{\textgreater{}}\NormalTok{       [,}\DecValTok{1}\NormalTok{] [,}\DecValTok{2}\NormalTok{] [,}\DecValTok{3}\NormalTok{] [,}\DecValTok{4}\NormalTok{] [,}\DecValTok{5}\NormalTok{]  [,}\DecValTok{6}\NormalTok{] [,}\DecValTok{7}\NormalTok{]    [,}\DecValTok{8}\NormalTok{]  [,}\DecValTok{9}\NormalTok{] [,}\DecValTok{10}\NormalTok{]}
\NormalTok{\%}\SpecialCharTok{\textgreater{}}\NormalTok{  [}\DecValTok{1}\NormalTok{,]    }\DecValTok{0}  \FloatTok{0.9}  \FloatTok{0.1}  \FloatTok{0.0} \FloatTok{0.00} \FloatTok{0.000}  \FloatTok{0.0} \FloatTok{0.00000} \FloatTok{0e+00} \FloatTok{0.000}
\NormalTok{\%}\SpecialCharTok{\textgreater{}}\NormalTok{  [}\DecValTok{2}\NormalTok{,]    }\DecValTok{0}  \FloatTok{1.0}  \FloatTok{0.0}  \FloatTok{0.0} \FloatTok{0.00} \FloatTok{0.000}  \FloatTok{0.0} \FloatTok{0.00000} \FloatTok{0e+00} \FloatTok{0.000}
\NormalTok{\%}\SpecialCharTok{\textgreater{}}\NormalTok{  [}\DecValTok{3}\NormalTok{,]    }\DecValTok{0}  \FloatTok{0.0}  \FloatTok{0.0}  \FloatTok{0.9} \FloatTok{0.09} \FloatTok{0.010}  \FloatTok{0.0} \FloatTok{0.00000} \FloatTok{0e+00} \FloatTok{0.000}
\NormalTok{\%}\SpecialCharTok{\textgreater{}}\NormalTok{  [}\DecValTok{4}\NormalTok{,]    }\DecValTok{0}  \FloatTok{0.0}  \FloatTok{0.0}  \FloatTok{0.0} \FloatTok{0.10} \FloatTok{0.001}  \FloatTok{0.0} \FloatTok{0.00000} \FloatTok{1e{-}01} \FloatTok{0.799}
\NormalTok{\%}\SpecialCharTok{\textgreater{}}\NormalTok{  [}\DecValTok{5}\NormalTok{,]    }\DecValTok{0}  \FloatTok{0.0}  \FloatTok{0.0}  \FloatTok{0.0} \FloatTok{0.00} \FloatTok{0.300}  \FloatTok{0.0} \FloatTok{0.60000} \FloatTok{1e{-}01} \FloatTok{0.000}
\NormalTok{\%}\SpecialCharTok{\textgreater{}}\NormalTok{  [}\DecValTok{6}\NormalTok{,]    }\DecValTok{0}  \FloatTok{0.0}  \FloatTok{0.0}  \FloatTok{0.0} \FloatTok{0.00} \FloatTok{0.000}  \FloatTok{0.7} \FloatTok{0.00000} \FloatTok{3e{-}01} \FloatTok{0.000}
\NormalTok{\%}\SpecialCharTok{\textgreater{}}\NormalTok{  [}\DecValTok{7}\NormalTok{,]    }\DecValTok{0}  \FloatTok{0.0}  \FloatTok{0.0}  \FloatTok{0.0} \FloatTok{0.00} \FloatTok{0.000}  \FloatTok{0.0} \FloatTok{0.32999} \FloatTok{1e{-}05} \FloatTok{0.670}
\NormalTok{\%}\SpecialCharTok{\textgreater{}}\NormalTok{  [}\DecValTok{8}\NormalTok{,]    }\DecValTok{0}  \FloatTok{0.0}  \FloatTok{0.0}  \FloatTok{0.0} \FloatTok{0.00} \FloatTok{0.000}  \FloatTok{0.0} \FloatTok{0.00000} \FloatTok{0e+00} \FloatTok{1.000}
\NormalTok{\%}\SpecialCharTok{\textgreater{}}\NormalTok{  [}\DecValTok{9}\NormalTok{,]    }\DecValTok{0}  \FloatTok{0.0}  \FloatTok{0.0}  \FloatTok{0.0} \FloatTok{0.00} \FloatTok{0.000}  \FloatTok{0.0} \FloatTok{0.00000} \FloatTok{1e+00} \FloatTok{0.000}
\NormalTok{\%}\SpecialCharTok{\textgreater{}}\NormalTok{ [}\DecValTok{10}\NormalTok{,]    }\DecValTok{0}  \FloatTok{0.0}  \FloatTok{0.0}  \FloatTok{0.0} \FloatTok{0.00} \FloatTok{0.000}  \FloatTok{0.0} \FloatTok{0.00000} \FloatTok{0e+00} \FloatTok{1.000}
\end{Highlighting}
\end{Shaded}

This matrix can be visualized as follows:

\begin{Shaded}
\begin{Highlighting}[]
\FunctionTok{visualizeGraph}\NormalTok{(}\AttributeTok{para=}\NormalTok{para, }\AttributeTok{option =} \StringTok{"P"}\NormalTok{)}
\end{Highlighting}
\end{Shaded}

\includegraphics{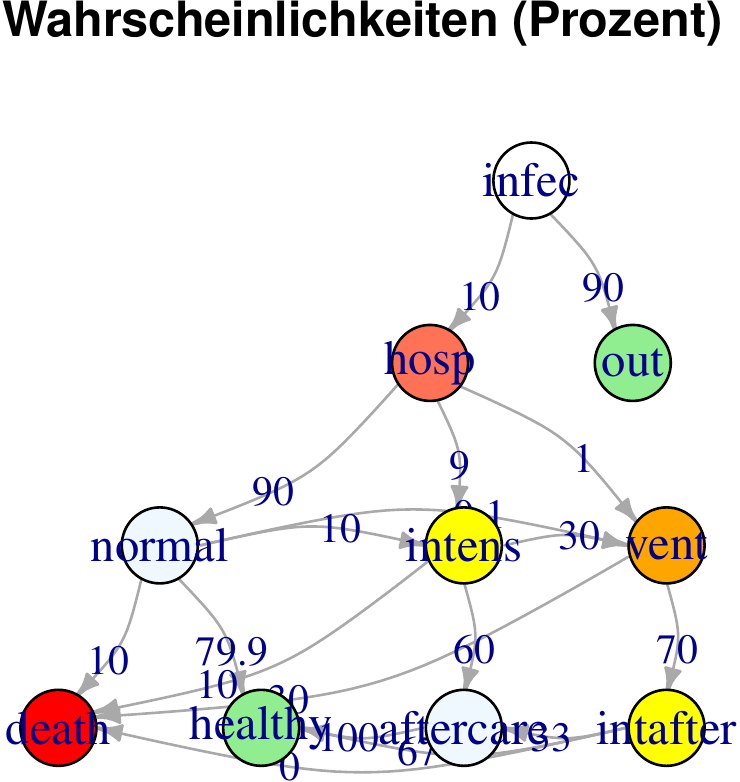}

Similar to the probabilities, durations can be visualized:

The corresponding matrix is shown below:

\hypertarget{extend-rki-data}{%
\section{Extend RKI Data}\label{extend-rki-data}}

\label{sec:extend}

\texttt{babsim.hospital}can be used to simulate several scenarios, i.e.,
possible developments of the pandemic. To simulate these scenarios,
arrival events must be generated. The function \texttt{extendRki()} adds
new arrival events. To generate new arrivals, three parameters must be
specified:

\begin{enumerate}
\item `data`: an already existing data set, i.e., the history
\item `EndDate`: last day of the simulated data (in the future)
\item `R0`: base reproduction values (R0) at the first day of the scenario and at the 
   last day of the scenario. A linear interpolation between these two values will 
   be used, e.g., if `R0 = c(1,2)` and ten eleven days are specified, the 
   following R0 values will be used: (1.0, 1.1, 1.2, 1.3, ..., 1.9,2.0).
\end{enumerate}

\begin{Shaded}
\begin{Highlighting}[]
\NormalTok{data }\OtherTok{\textless{}{-}} \FunctionTok{getRkiData}\NormalTok{(babsim.hospital}\SpecialCharTok{::}\NormalTok{rkidata)}
\NormalTok{\%}\SpecialCharTok{\textgreater{}} \FunctionTok{getRkiData}\NormalTok{()}\SpecialCharTok{:}\NormalTok{ Found days with negative number of cases. Ignoring them.}
\NormalTok{n }\OtherTok{\textless{}{-}}  \FunctionTok{as.integer}\NormalTok{( }\FunctionTok{max}\NormalTok{(data}\SpecialCharTok{$}\NormalTok{Day)}\SpecialCharTok{{-}}\FunctionTok{min}\NormalTok{(data}\SpecialCharTok{$}\NormalTok{Day) )}
\NormalTok{StartDay }\OtherTok{\textless{}{-}} \FunctionTok{min}\NormalTok{(data}\SpecialCharTok{$}\NormalTok{Day) }\SpecialCharTok{+} \FunctionTok{round}\NormalTok{(n}\SpecialCharTok{*}\FloatTok{0.9}\NormalTok{)  }
\NormalTok{data }\OtherTok{\textless{}{-}}\NormalTok{ data[}\FunctionTok{which}\NormalTok{(data}\SpecialCharTok{$}\NormalTok{Day }\SpecialCharTok{\textgreater{}=}\NormalTok{  StartDay), ]}
\NormalTok{EndDate }\OtherTok{\textless{}{-}} \FunctionTok{max}\NormalTok{(data}\SpecialCharTok{$}\NormalTok{Day) }\SpecialCharTok{+} \DecValTok{14}
\NormalTok{dataExt }\OtherTok{\textless{}{-}} \FunctionTok{extendRki}\NormalTok{(}\AttributeTok{data =}\NormalTok{ data, }
                     \AttributeTok{EndDate =}\NormalTok{ EndDate,}
                     \AttributeTok{R0 =} \FunctionTok{c}\NormalTok{(}\FloatTok{0.1}\NormalTok{, }\FloatTok{0.2}\NormalTok{))}
\end{Highlighting}
\end{Shaded}

To illustrate the \texttt{extendRki()} data extension procedure, a short
example is shown below:

\begin{Shaded}
\begin{Highlighting}[]
\FunctionTok{visualizeRkiEvents}\NormalTok{(}\AttributeTok{data =}\NormalTok{ data, }\AttributeTok{region=}\DecValTok{5374}\NormalTok{)}
\end{Highlighting}
\end{Shaded}

\includegraphics{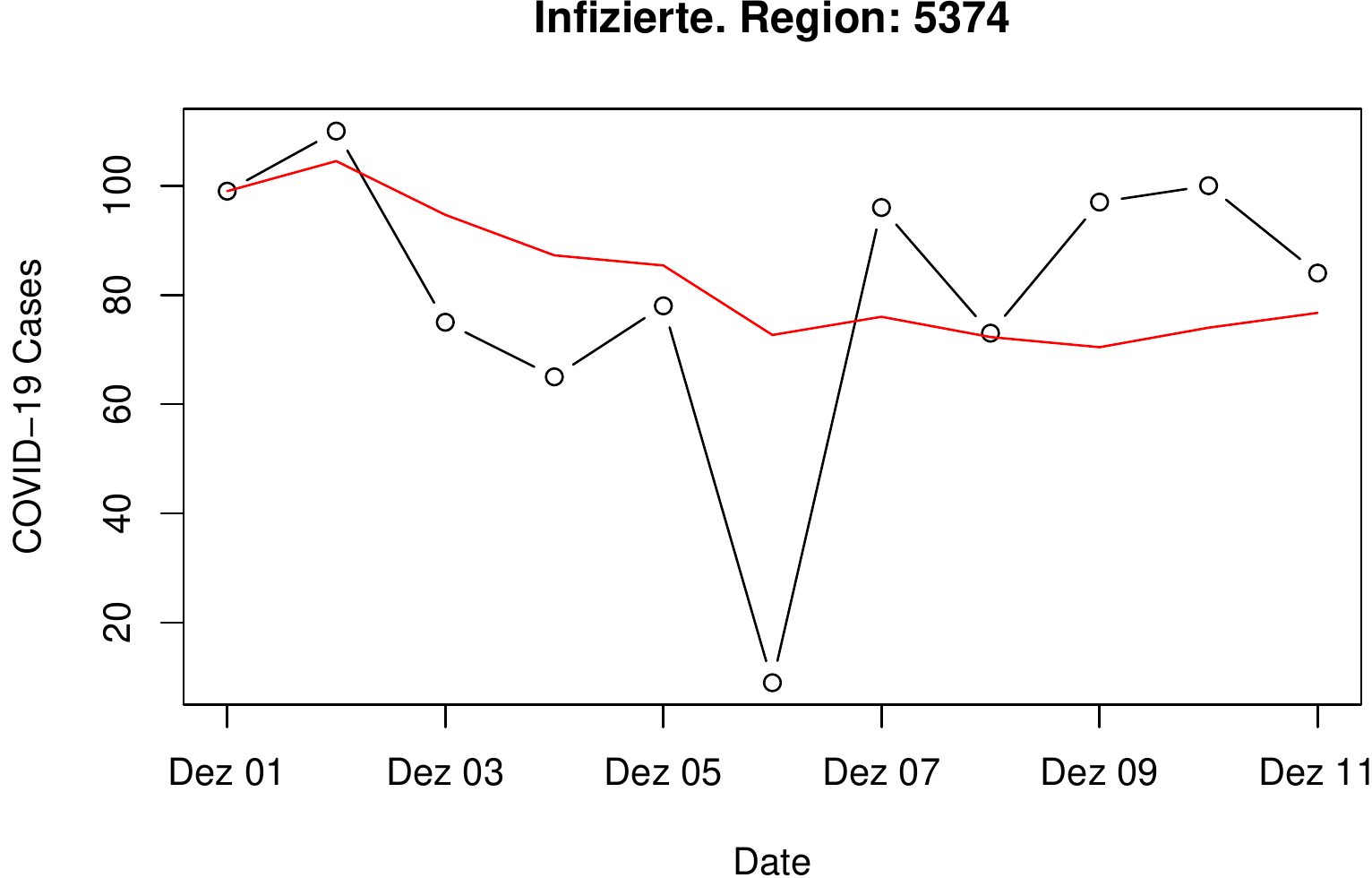}

The following plot shows the result of the data extension:

\begin{Shaded}
\begin{Highlighting}[]
\FunctionTok{visualizeRkiEvents}\NormalTok{(}\AttributeTok{data =}\NormalTok{ dataExt, }\AttributeTok{region =} \DecValTok{5374}\NormalTok{)}
\end{Highlighting}
\end{Shaded}

\includegraphics{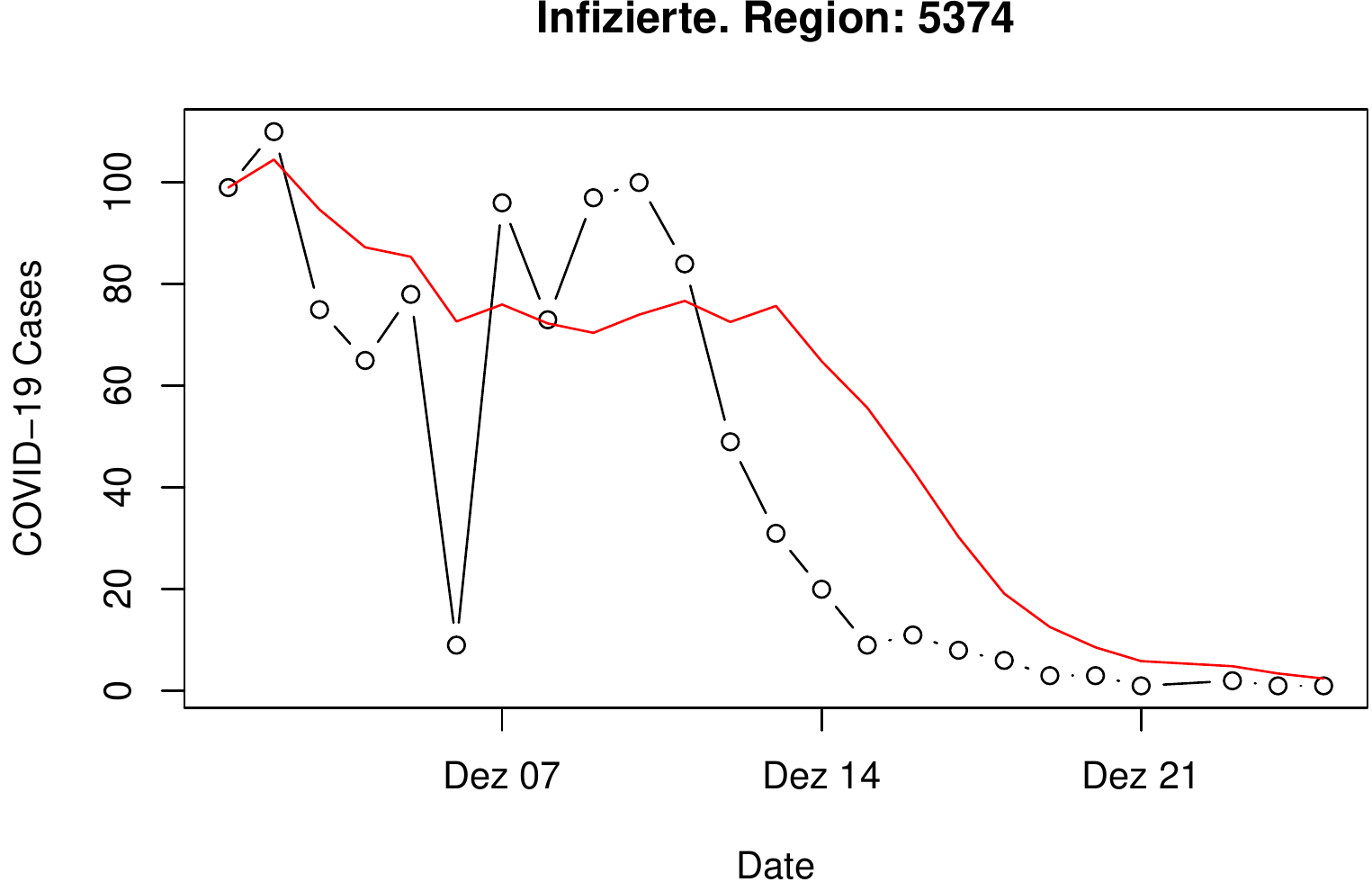}

\hypertarget{sensitivity-analysis}{%
\section{Sensitivity Analysis}\label{sensitivity-analysis}}

\label{sec:sensitivity}

We will describe only a very quick parameter analysis. A detailed
analysis will be presented in a forthcoming paper. Here, we demonstrate
how machine laerning tools (regression trees) can be used to visualize
the most important parameters of the simulation model.

\begin{Shaded}
\begin{Highlighting}[]
\NormalTok{param }\OtherTok{\textless{}{-}} \FunctionTok{getParaSet}\NormalTok{(}\DecValTok{5374}\NormalTok{)}
\NormalTok{n }\OtherTok{\textless{}{-}} \FunctionTok{dim}\NormalTok{(param)[}\DecValTok{2}\NormalTok{] }\SpecialCharTok{{-}} \DecValTok{1}
\NormalTok{y }\OtherTok{\textless{}{-}}\NormalTok{ param[,}\DecValTok{1}\NormalTok{]}
\NormalTok{x }\OtherTok{\textless{}{-}}\NormalTok{ param[,}\DecValTok{2}\SpecialCharTok{:}\FunctionTok{dim}\NormalTok{(param)[}\DecValTok{2}\NormalTok{]]}
\NormalTok{fitTree }\OtherTok{\textless{}{-}} \FunctionTok{buildTreeModel}\NormalTok{(}\AttributeTok{x=}\NormalTok{x,}
                 \AttributeTok{y=}\NormalTok{y,}
                 \AttributeTok{control =} \FunctionTok{list}\NormalTok{(}\AttributeTok{xnames =} \FunctionTok{paste0}\NormalTok{(}\StringTok{\textquotesingle{}x\textquotesingle{}}\NormalTok{, }\DecValTok{1}\SpecialCharTok{:}\NormalTok{n)))}
\FunctionTok{rpart.plot}\NormalTok{(fitTree}\SpecialCharTok{$}\NormalTok{fit)}
\end{Highlighting}
\end{Shaded}

\includegraphics{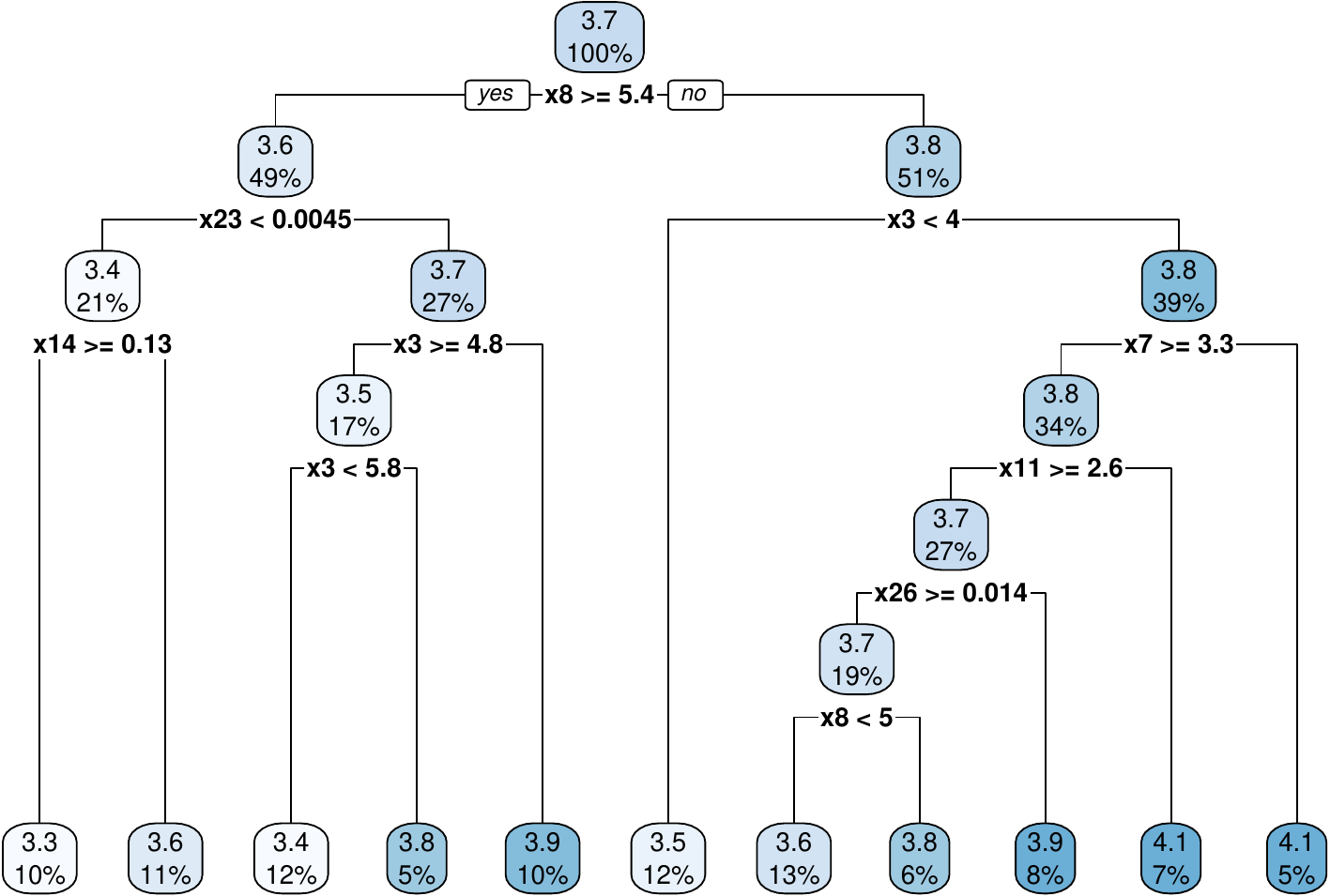}

\begin{Shaded}
\begin{Highlighting}[]
\FunctionTok{getParameterNameList}\NormalTok{(}\FunctionTok{c}\NormalTok{(}\DecValTok{24}\NormalTok{, }\DecValTok{25}\NormalTok{, }\DecValTok{3}\NormalTok{, }\DecValTok{10}\NormalTok{))}
\NormalTok{\%}\SpecialCharTok{\textgreater{}}\NormalTok{                          x24                          x25 }
\NormalTok{\%}\SpecialCharTok{\textgreater{}} \StringTok{"AmntDaysAftercareToHealthy"}                \StringTok{"RiskFactorA"} 
\NormalTok{\%}\SpecialCharTok{\textgreater{}}\NormalTok{                           x3                          x10 }
\NormalTok{\%}\SpecialCharTok{\textgreater{}}  \StringTok{"AmntDaysNormalToIntensive"} \StringTok{"AmntDaysVentilationToDeath"}
\end{Highlighting}
\end{Shaded}

Sevral additional tools are available, e.g., to perform a sensitivity
analysis, because \texttt{babsim.hopital} uses the \texttt{R} package
\texttt{SPOT} (sequential parameter optimization toolbox) to improve
parameter settings.\texttt{SPOT} implements a set of tools for
model-based optimization and tuning of algorithms (surrogate models,
optimizers, DOE). \texttt{SPOT} can be used for sensitivity analysis,
which is in important under many aspects, especially:

\begin{itemize}
\item  understanding the most important factors (parameters) that influence model
   behavior. For example, it is of great importance for simulation practitioners and
   doctors to discover relevant durations and probabilities. 
\item detecting interactions between parameters, e.g., do durations influence 
   each other?
\end{itemize}

The fitness landscape can be visualized using the function
\texttt{plotModel}. Again, we would like to mention that these results
are only shown for didactical purposes. They are not a valid statistical
analysis, because the optimization via simulation runs are too short and
do not produce sufficient results. Here, the interaction between the
first two model parameters, i.e., AmntDaysInfectedToHospital, and
AmntDaysNormalToHealthy, is shown.

\begin{Shaded}
\begin{Highlighting}[]
\NormalTok{SPOT}\SpecialCharTok{::}\FunctionTok{plotModel}\NormalTok{(res}\SpecialCharTok{$}\NormalTok{modelFit, }\AttributeTok{which =} \FunctionTok{c}\NormalTok{(}\DecValTok{1}\NormalTok{,}\DecValTok{2}\NormalTok{) ,}\AttributeTok{xlab =} \FunctionTok{c}\NormalTok{(}\StringTok{"x1"}\NormalTok{, }\StringTok{"x2"}\NormalTok{))}
\end{Highlighting}
\end{Shaded}

\includegraphics{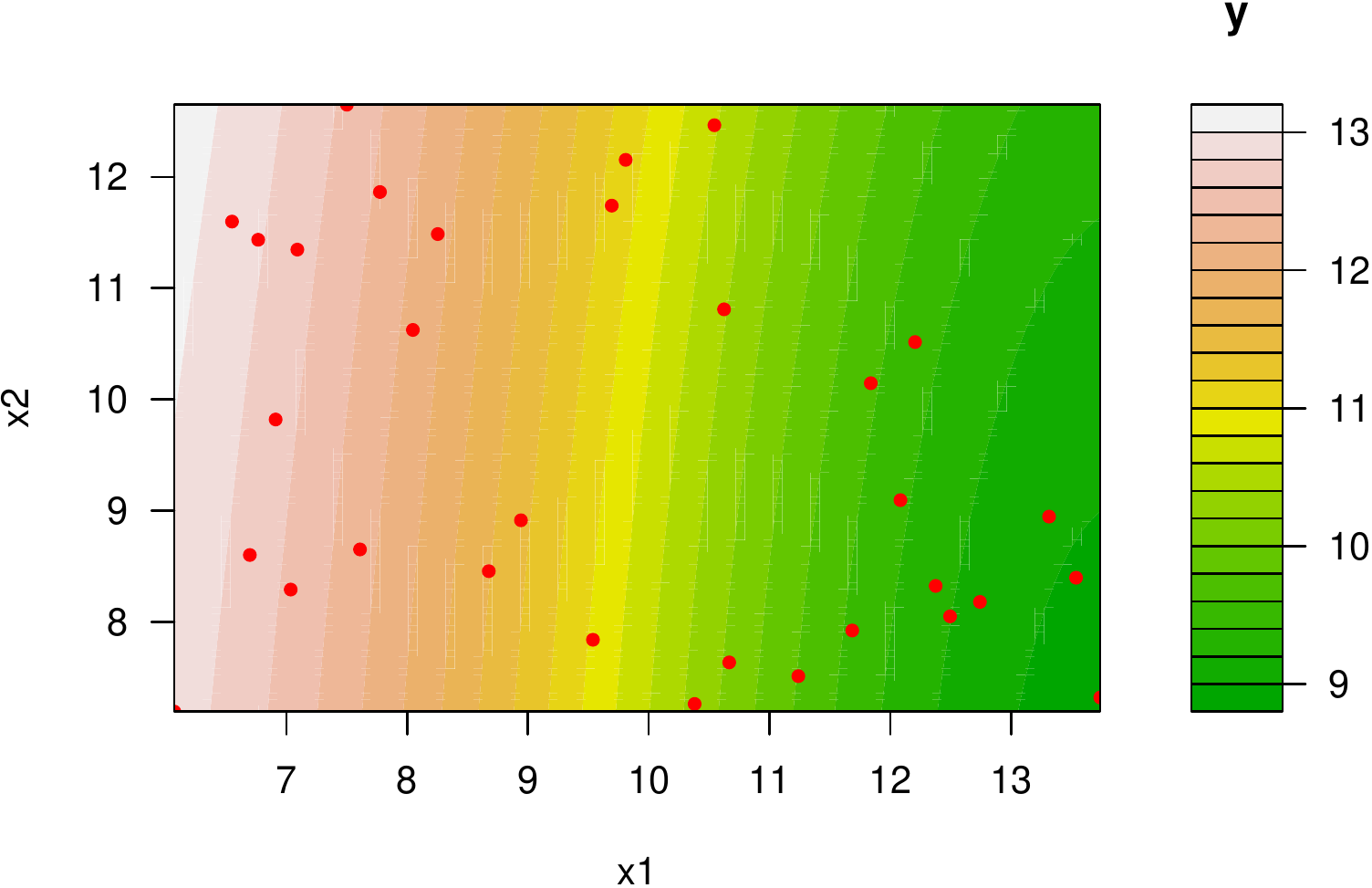} A
regression-based parameter screening can be performed to discover
relevant (and irrelevant) model parameters:

\begin{Shaded}
\begin{Highlighting}[]
\NormalTok{fitLm }\OtherTok{\textless{}{-}}\NormalTok{ SPOT}\SpecialCharTok{::}\FunctionTok{buildLM}\NormalTok{(}\AttributeTok{x=}\NormalTok{res}\SpecialCharTok{$}\NormalTok{x,}
                 \AttributeTok{y=}\NormalTok{res}\SpecialCharTok{$}\NormalTok{y,}
                 \AttributeTok{control =} \FunctionTok{list}\NormalTok{(}\AttributeTok{useStep=}\ConstantTok{TRUE}\NormalTok{))}
\FunctionTok{summary}\NormalTok{(fitLm}\SpecialCharTok{$}\NormalTok{fit)}
\NormalTok{\%}\SpecialCharTok{\textgreater{}} 
\NormalTok{\%}\SpecialCharTok{\textgreater{}}\NormalTok{ Call}\SpecialCharTok{:}
\NormalTok{\%}\SpecialCharTok{\textgreater{}} \FunctionTok{lm}\NormalTok{(}\AttributeTok{formula =}\NormalTok{ y }\SpecialCharTok{\textasciitilde{}}\NormalTok{ x}\FloatTok{.12} \SpecialCharTok{+}\NormalTok{ x}\FloatTok{.13} \SpecialCharTok{+}\NormalTok{ x}\FloatTok{.14} \SpecialCharTok{+}\NormalTok{ x}\FloatTok{.15} \SpecialCharTok{+}\NormalTok{ x}\FloatTok{.24} \SpecialCharTok{+}\NormalTok{ x}\FloatTok{.25} \SpecialCharTok{+}\NormalTok{ x}\FloatTok{.26}\NormalTok{, }
\NormalTok{\%}\SpecialCharTok{\textgreater{}}     \AttributeTok{data =}\NormalTok{ df)}
\NormalTok{\%}\SpecialCharTok{\textgreater{}} 
\NormalTok{\%}\SpecialCharTok{\textgreater{}}\NormalTok{ Residuals}\SpecialCharTok{:}
\NormalTok{\%}\SpecialCharTok{\textgreater{}}\NormalTok{     Min      1Q  Median      3Q     Max }
\NormalTok{\%}\SpecialCharTok{\textgreater{}} \SpecialCharTok{{-}}\FloatTok{3.7513} \SpecialCharTok{{-}}\FloatTok{1.2679} \SpecialCharTok{{-}}\FloatTok{0.1442}  \FloatTok{1.0313}  \FloatTok{6.1996} 
\NormalTok{\%}\SpecialCharTok{\textgreater{}} 
\NormalTok{\%}\SpecialCharTok{\textgreater{}}\NormalTok{ Coefficients}\SpecialCharTok{:}
\NormalTok{\%}\SpecialCharTok{\textgreater{}}\NormalTok{             Estimate Std. Error t value }\FunctionTok{Pr}\NormalTok{(}\SpecialCharTok{\textgreater{}}\ErrorTok{|}\NormalTok{t}\SpecialCharTok{|}\NormalTok{)    }
\NormalTok{\%}\SpecialCharTok{\textgreater{}}\NormalTok{ (Intercept)  }\FloatTok{16.9509}     \FloatTok{2.0349}   \FloatTok{8.330} \FloatTok{2.14e{-}13} \SpecialCharTok{**}\ErrorTok{*}
\NormalTok{\%}\SpecialCharTok{\textgreater{}}\NormalTok{ x}\FloatTok{.12}          \FloatTok{0.2486}     \FloatTok{0.1004}   \FloatTok{2.475}   \FloatTok{0.0148} \SpecialCharTok{*}  
\NormalTok{\%}\SpecialCharTok{\textgreater{}}\NormalTok{ x}\FloatTok{.13}          \FloatTok{0.5088}     \FloatTok{0.3603}   \FloatTok{1.412}   \FloatTok{0.1606}    
\NormalTok{\%}\SpecialCharTok{\textgreater{}}\NormalTok{ x}\FloatTok{.14}        \SpecialCharTok{{-}}\FloatTok{31.8801}     \FloatTok{5.6697}  \SpecialCharTok{{-}}\FloatTok{5.623} \FloatTok{1.38e{-}07} \SpecialCharTok{**}\ErrorTok{*}
\NormalTok{\%}\SpecialCharTok{\textgreater{}}\NormalTok{ x}\FloatTok{.15}        \SpecialCharTok{{-}}\FloatTok{30.2948}    \FloatTok{14.6745}  \SpecialCharTok{{-}}\FloatTok{2.064}   \FloatTok{0.0413} \SpecialCharTok{*}  
\NormalTok{\%}\SpecialCharTok{\textgreater{}}\NormalTok{ x}\FloatTok{.24}         \SpecialCharTok{{-}}\FloatTok{0.4782}     \FloatTok{0.2884}  \SpecialCharTok{{-}}\FloatTok{1.658}   \FloatTok{0.1000}    
\NormalTok{\%}\SpecialCharTok{\textgreater{}}\NormalTok{ x}\FloatTok{.25}          \FloatTok{1.5502}     \FloatTok{0.5214}   \FloatTok{2.973}   \FloatTok{0.0036} \SpecialCharTok{**} 
\NormalTok{\%}\SpecialCharTok{\textgreater{}}\NormalTok{ x}\FloatTok{.26}         \FloatTok{22.1438}     \FloatTok{9.3273}   \FloatTok{2.374}   \FloatTok{0.0193} \SpecialCharTok{*}  
\NormalTok{\%}\SpecialCharTok{\textgreater{}} \SpecialCharTok{{-}{-}{-}}
\NormalTok{\%}\SpecialCharTok{\textgreater{}}\NormalTok{ Signif. codes}\SpecialCharTok{:}  \DecValTok{0} \StringTok{\textquotesingle{}***\textquotesingle{}} \FloatTok{0.001} \StringTok{\textquotesingle{}**\textquotesingle{}} \FloatTok{0.01} \StringTok{\textquotesingle{}*\textquotesingle{}} \FloatTok{0.05} \StringTok{\textquotesingle{}.\textquotesingle{}} \FloatTok{0.1} \StringTok{\textquotesingle{} \textquotesingle{}} \DecValTok{1}
\NormalTok{\%}\SpecialCharTok{\textgreater{}} 
\NormalTok{\%}\SpecialCharTok{\textgreater{}}\NormalTok{ Residual standard error}\SpecialCharTok{:} \FloatTok{1.785}\NormalTok{ on }\DecValTok{113}\NormalTok{ degrees of freedom}
\NormalTok{\%}\SpecialCharTok{\textgreater{}}\NormalTok{   (}\DecValTok{1}\NormalTok{ observation deleted due to missingness)}
\NormalTok{\%}\SpecialCharTok{\textgreater{}}\NormalTok{ Multiple R}\SpecialCharTok{{-}}\NormalTok{squared}\SpecialCharTok{:}  \FloatTok{0.3807}\NormalTok{, Adjusted R}\SpecialCharTok{{-}}\NormalTok{squared}\SpecialCharTok{:}  \FloatTok{0.3423} 
\NormalTok{\%}\SpecialCharTok{\textgreater{}}\NormalTok{ F}\SpecialCharTok{{-}}\NormalTok{statistic}\SpecialCharTok{:} \FloatTok{9.924}\NormalTok{ on }\DecValTok{7}\NormalTok{ and }\DecValTok{113}\NormalTok{ DF,  p}\SpecialCharTok{{-}}\NormalTok{value}\SpecialCharTok{:} \FloatTok{1.307e{-}09}
\end{Highlighting}
\end{Shaded}

\hypertarget{summary}{%
\section{Summary}\label{summary}}

Based on ideas presented by Lawton and McCooe (2019), we developed a
resource planning tool for hospitals that considers the specific
situation of the COVID-19 pandemic. \texttt{babsim.hospital} is
implemented in the statistical programming language R, see R Core Team
(2020), and uses a discrete-event simulation model. The \texttt{simmer}
package (Ucar, Smeets, and Azcorra 2019) was used to implement the
simulation model.

\texttt{babsimhospital} was developed in cooperation with ICU experts,
crisis teams, and health administrations. It combines simulation,
optimization, statistics, and artificial intelligence processes in a
very efficient way.

Sequential parameter optimization (SPOT) is used to optimize the
parameters (status transition probabilities, length of stay,
distribution properties) Bartz-Beielstein, Lasarczyk, and Preuss (2005).
For modeling purposes, distributions (including a truncated and
translated Gamma distribution) were specially developed by us in order
to realistically simulate the length of stay. babsim.hospital takes into
account different risks for individual groups of people (age and
gender-specific) and can be used for any resources beyond the planning
of bed capacities.

Using \texttt{babsim.hospital} provides many advantages for crisis
teams, e.g., comparison with your own local planning, simulation of
local events, adaptation to your own situation, simulation of any
scenario (worst / best case), simulation of any pandemic scenario,
considering the local situation, and enables a standardized approach.
And, there are benefits for medical professionals, e.g, analysis of the
pandemic at local, regional, state and federal level, the consideration
of special risk groups, tools for validating the length of stays and
transition probabilities. Finally, there are potential advantages for
administration, management, e.g., assessment of the situation of
individual hospitals taking local events into account, consideration of
relevant resources: beds, ventilators, rooms, protective clothing, and
personnel planning, e.g., medical and nursing staff.

\texttt{babsim.hospital} is open source and will be available on CRAN,
see (\url{https://cran.r-project.org}).

\hypertarget{appendix}{%
\section{Appendix}\label{appendix}}

\label{sec:appendix}

\hypertarget{copyright-notices}{%
\subsection{Copyright Notices}\label{copyright-notices}}

\hypertarget{rki-data}{%
\subsubsection{RKI Data}\label{rki-data}}

Please take the following copyright notice under advisement, if you plan
to use the RKI data included in the package:

\begin{quote}
Die Daten sind die „Fallzahlen in Deutschland`` des Robert Koch-Institut
(RKI) und stehen unter der Open Data Datenlizenz Deutschland Version 2.0
zur Verfügung. Quellenvermerk: Robert Koch-Institut (RKI),
dl-de/by-2-0\\
Haftungsausschluss: „Die Inhalte, die über die Internetseiten des Robert
Koch-Instituts zur Verfügung gestellt werden, dienen ausschließlich der
allgemeinen Information der Öffentlichkeit, vorrangig der
Fachöffentlichkeit``.
\end{quote}

Taken from
\url{https://npgeo-corona-npgeo-de.hub.arcgis.com/datasets/dd4580c810204019a7b8eb3e0b329dd6_0}.

\hypertarget{divi-data}{%
\subsubsection{DIVI Data}\label{divi-data}}

We have included a data sample from the German
\href{https://www.intensivregister.de/}{DIVI Register:
https://www.intensivregister.de/}. Please take the following copyright
notice under advisement. The DIVI data are not open data. The following
statement can be found on the DIVI web page:

\begin{quote}
Eine weitere wissenschaftliche Nutzung der Daten ist nur mit Zustimmung
der DIVI gestattet. Therefore, only an example data set, that reflects
the structure of the original data from the DIVI register, is included
in the \texttt{babsim.hospital} package as \texttt{icudata}.
\end{quote}

\hypertarget{references}{%
\section{References}\label{references}}

\hypertarget{refs}{}
\begin{CSLReferences}{1}{0}
\leavevmode\hypertarget{ref-Bart20j}{}%
Bartz-Beielstein, Thomas, Eva Bartz, Frederik Rehbach, and Olaf
Mersmann. 2020. {``{Optimization of High-dimensional Simulation Models
Using Synthetic Data}.''} \emph{arXiv e-Prints}, September,
arXiv:2009.02781. \url{http://arxiv.org/abs/2009.02781}.

\leavevmode\hypertarget{ref-BLP05}{}%
Bartz-Beielstein, Thomas, Christian Lasarczyk, and Mike Preuss. 2005.
{``{Sequential Parameter Optimization}.''} In \emph{Proceedings 2005
Congress on Evolutionary Computation (CEC'05), Edinburgh, Scotland},
edited by B McKay and others, 773--80. Piscataway NJ: IEEE Press.
\url{https://doi.org/10.1109/CEC.2005.1554761}.

\leavevmode\hypertarget{ref-BLP05}{}%
---------. 2005. {``{Sequential Parameter Optimization}.''} In
\emph{Proceedings 2005 Congress on Evolutionary Computation (CEC'05),
Edinburgh, Scotland}, edited by B McKay and others, 773--80. Piscataway
NJ: IEEE Press. \url{https://doi.org/10.1109/CEC.2005.1554761}.

\leavevmode\hypertarget{ref-Lawt19a}{}%
Lawton, Tom, and Michael McCooe. 2019. {``Policy: A Novel Modelling
Technique to Predict Resource -Requirements in Critical Care {{}} a Case
Study.''} \emph{Future Healthcare Journal} 6 (1): 17--20.
\url{https://doi.org/10.7861/futurehosp.6-1-17}.

\leavevmode\hypertarget{ref-R20a}{}%
R Core Team. 2020. \emph{R: A Language and Environment for Statistical
Computing}. Vienna, Austria: R Foundation for Statistical Computing.
\url{https://www.R-project.org/}.

\leavevmode\hypertarget{ref-Ucar19a}{}%
Ucar, Iñaki, Bart Smeets, and Arturo Azcorra. 2019. {``Simmer:
Discrete-Event Simulation for r.''} \emph{Journal of Statistical
Software, Articles} 90 (2): 1--30.
\url{https://doi.org/10.18637/jss.v090.i02}.

\end{CSLReferences}

\end{document}